\documentclass{article}
\usepackage[active]{srcltx}
\usepackage{amsmath}
\usepackage{amsfonts}
\usepackage{amssymb}
\usepackage{amscd}
\usepackage{a4wide}
\usepackage{bm}
\usepackage{amssymb}
\usepackage{latexsym}
\usepackage{amsmath}
\usepackage{amsfonts}
\usepackage{color}
\usepackage{dsfont}
\usepackage{indentfirst}
\usepackage[latin1]{inputenc}
\usepackage{mathrsfs}
\usepackage{bbm}
\usepackage{yfonts}

\def\qed{\hfill $\square$}

\newtheorem{definition}{\underline{Definition}}[section]
\newtheorem{proposition}[definition]{Proposition}
\newtheorem{theorem}[definition]{Theorem}

\newtheorem{remark}[definition]{Remark}
\newtheorem{lema}[definition]{Lemma}

\numberwithin{equation}{section}

\DeclareMathAlphabet{\mathpzc}{OT1}{pzc}{m}{it}

\begin{document}
\begin{center}
{\Large{ \textbf{On the optical properties of carbon nanotubes--Part I.\\ A general formula for the dynamical optical conductivity.}}}

\medskip

\today
\end{center}

\begin{center}
\small{Morten Grud Rasmussen\footnote{Department of Mathematical Sciences, Aalborg University, Fredrik Bajers Vej 7G, 9220 Aalborg, Denmark; e-mail: morteng@math.aau.dk}, Benjamin Ricaud\footnote{Laboratoire de Traitement des Signaux 2, \'Ecole Polytechnique F\'ed\'erale de Lausanne,
Lausanne, Vaud, Switzerland; e-mail: benjamin.ricaud@epfl.ch}, Baptiste Savoie\footnote{Dublin Institute for Advanced Studies, School of Theoretical Physics, 10 Burlington road, Dublin 04, Ireland; e-mail: baptiste.savoie@gmail.com}}
\end{center}
\vspace{1.0cm}

\begin{abstract}
This paper is the first one in a series of two articles in which we revisit the optical properties of single-walled carbon nanotubes (SWNT). Produced by rolling up a graphene sheet, SWNT owe their intriguing properties to their cylindrical quasi-one-dimensional (quasi-1D) structure (the ratio length/radius is experimentally  of order of $10^{3}$). We model SWNT by circular cylinders of small diameters on the surface of which the conduction electron gas is confined by the electric field generated by the fixed carbon ions. The pair-interaction potential considered is the 3D Coulomb potential restricted to the cylinder. To reflect the quasi-1D structure, we introduce a 1D effective many-body Hamiltonian which is the starting-point of our analysis. To investigate the optical properties, we consider a perturbation by a uniform time-dependent electric field modeling an incident light beam along the longitudinal direction. By using Kubo's method, we derive within the linear response theory an asymptotic expansion in the low-temperature regime for the dynamical optical conductivity at fixed density of particles. The leading term only involves the eigenvalues and associated eigenfunctions of the (unperturbed) 1D effective many-body Hamiltonian, and allows us to account for the sharp peaks observed in the optical absorption spectrum of SWNT.
\end{abstract}
\vspace{1.0cm}

\noindent
\textbf{PACS-2010 number}: 78.67.Ch, 78.67.-n, 73.63.Fg, 71.10.Ca.

\medskip

\noindent
\textbf{MSC-2010 number}: 82C22, 82D80, 82D37, 82B10, 81Q10.	

\medskip

\medskip

\tableofcontents
\medskip
\vspace{1.0cm}

\section{The settings and the main result.}

\subsection{Modeling SWNT as quasi-one dimensional structures.}
\label{modl}

\textbf{General assumptions--Infinite volume systems.} Consider a circular cylinder of infinite length with radius $r>0$ on the surface of which lies a regular lattice of carbon atoms. We suppose that the lattice is fixed 
and forms a non-degenerate periodic pattern of hexagons. Denote by $a$ and $b=b(r)$ respectively the longitudinal and transverse periods of the lattice. Due to the configuration, there exists $l \in \mathbb{N}^{*}$ s.t. $l b=2\pi r$. Besides, we assume that \textit{only one electron per carbon atom} is likely to be delocalized, and then plays the role of conduction electron. Each conduction electron is confined on the surface of the cylinder by the electric field generated by the positive carbon ions. The associated electric potential energy $V_{\mathrm{per}}$ is supposed to be periodic w.r.t. the lattice and uniformly locally square-integrable, i.e. $V_{\mathrm{per}} \in L^{2}_{\mathrm{uloc}}(\mathbb{R}^{2})$ see e.g.
\cite[Sec. XIII.16]{RS4}. In particular, this implies that $V_{\mathrm{per}}$ is square-integrable on the unit cell of the lattice. We point out that $V_{\mathrm{per}}$ contains all the information about the chirality (i.e. 'twist') of the tube. Furthermore, the conduction electrons interact with each other. The pair-interaction potential energy that we consider is the 3D Coulomb potential restricted to the cylinder:
\begin{equation}
\label{Vcrdef}
V_{r}(x,y) := \frac{e^{2}}{\varepsilon \sqrt{x^{2} + 4 r^{2} \sin^{2}(\frac{y}{2r})}}, \quad (x,y) \in \mathbb{R}\times r\mathbb{S},
\end{equation}
where $e$ denotes the elementary charge and $\varepsilon$ the electric permittivity of the material assumed to be constant. Hereafter, we denote by $\mathcal{C}_{\infty,r}:= \mathbb{R}\times r\mathbb{S}$ the cylinder surface where $\mathbb{S}:=\mathbb{R}/(2\pi\mathbb{Z})$ stands for the unit circle. \eqref{Vcrdef} is justified by Pythagora's theorem. The cylinder is embedded in $\mathbb{R}^{3}$. The distance $\rho$ from one particle to the other  in $\mathbb{R}^{3}$ reads as $\rho^{2}=(x_{1}-x_{2})^{2} + 4r^{2} \sin^{2}(\frac{y_{1} - y_{2}}{2r})$, where $\vert 2r \sin(\frac{y_{1} - y_{2}}{2r})\vert$ is the length of the chord joining two points of coordinate $y_{1}$ and $y_{2}$ on the circle. From \eqref{Vcrdef}, $V_{r} \in L_{\mathrm{loc}}^{1}(\mathcal{C}_{\infty,r})$  but $V_{r}(\cdot\,,y) \notin L^{1}(\mathbb{R})$ even for $y \neq 0$. Nevertheless, its Fourier transform exists whenever $y\neq 0$ and it has the explicit expression:
\begin{equation}
\label{TFVcr}
\forall p \in \mathbb{R}^{*},\quad \widehat{V_{r}}(p,y) := \frac{1}{\sqrt{2\pi}} \int_{\mathbb{R}} \mathrm{d}x'\, \mathrm{e}^{- i p x'} V_{r}(x',y) = \sqrt{\frac{2}{\pi}} \frac{e^{2}}{\varepsilon} K_{0}( 2 r \vert p \sin(\frac{y}{2r})\vert),
\end{equation}
where $K_{0}$ is the modified Bessel function of the second kind, see \cite{AS}. Note that $K_{0} \in L^{1}([0,\infty))$. We refer the readers to \cite{CDR1} for a spectral analysis of the Hamiltonian describing the dynamics of two self-interacting charges of opposite sign (the so-called 'exciton model') confined on $\mathcal{C}_{\infty,r}$. The pair-interaction potential energy considered is precisely \eqref{Vcrdef}. \\
\indent \textbf{Many-body Hamiltonian for finite volume systems}. Let $\Lambda \times r\mathbb{S}$ be a strict subset of $\mathcal{C}_{\infty,r}$, where $\Lambda$ is a non-empty interval centered at the origin of coordinates. For convenience, we take $\Lambda = [-La/2,La/2)$ with $L \in \mathbb{N}^{*}$ so that its Lebesgue measure satisfies: $\vert \Lambda\vert = L a$, $L \in \mathbb{N}^{*}$. If  $N\in \mathbb{N}^{*}$ carbon ions lie on $\Lambda\times r\mathbb{S}$, then our assumptions imply that the number of conduction electrons is $N$. Note that the parameters $L,N,a,b$ are interrelated  since the tube length and its perimeter are multiples of an integer number of ions. Indeed, for $\mathpzc{n}_{0}$ ions in the polygon of area  $a \times b$:
\begin{equation}
\label{relsemic}
\frac{L a}{a} \times \frac{2\pi r}{b} \times \mathpzc{n}_{0} = N.
\end{equation}
A natural approach to investigate the dynamics of conduction electrons consists in imposing Dirichlet boundary conditions on $\Lambda$ (longitudinal direction) when defining the many-body Hamiltonian. This will however break its translational invariance. The standard approach instead consists in imposing periodic boundary conditions on $\Lambda$; this boils down to working on the torus $\mathbb{T}_{La}\times r\mathbb{S}$. Here and hereafter, we identify $\tau$-periodic functions on $\mathbb{R}$ with functions on the 1-dimensional torus: $\mathbb{T}_{\tau} := \mathbb{R}/(\tau \mathbb{Z})$ which we define by identifying points in $\mathbb{R}$ that differ by $\tau n$ for some $n \in \mathbb{Z}$. From now on, we assume that $0<2\sqrt{2}r<a$ without loss of generality.
Denote by $\mathcal{T}_{La,r}:= \mathbb{T}_{La}\times r\mathbb{S}$ the one-electron configuration space. Working on $\mathcal{T}_{La,r}$ requires to introduce a periodization (with period $La$) of $V_{r}(\cdot\,,y)$ in \eqref{Vcrdef} that converges point-wise to $V_{r}$ when $L \rightarrow \infty$. From the Poisson summation formula and \eqref{TFVcr}, we may suggest the following $La$-periodic symmetric function:
\begin{equation}
\label{essa}
\mathbb{R} \owns x \mapsto \frac{\sqrt{2\pi}}{La} \sum_{m\in \mathbb{Z}^{*}} \mathrm{e}^{i \frac{2\pi}{La} m x} \widehat{V_{r}}(\frac{2\pi}{La}m, y),\quad y \neq 0.
\end{equation}
We discarded the mode $m=0$ since $V_{r}(\cdot\,,y)\notin L^{1}(\mathbb{R})$. Note that \eqref{essa} can be rewritten as follows:
\begin{multline*}
\frac{1}{La} \sum_{m \in \mathbb{Z}} \mathrm{e}^{i \frac{2\pi}{La} m x} \int_{-\frac{La}{2}}^{\frac{La}{2}} \mathrm{d}x'\, \mathrm{e}^{- i \frac{2\pi}{La} m x'}V_{r}(x',y) \\
- \frac{1}{La} \int_{-\frac{La}{2}}^{\frac{La}{2}} \mathrm{d}x'\, V_{r}(x',y) +
 \frac{1}{La} \sum_{m \in \mathbb{Z}^{*}} \mathrm{e}^{i \frac{2\pi}{La} m x} \int_{\vert x'\vert \geq \frac{La}{2}} \mathrm{d}x'\, \mathrm{e}^{- i \frac{2\pi}{La} m x'}V_{r}(x',y).
\end{multline*}
The first term corresponds to the complete Fourier periodic expansion of $V_{r}(\cdot\,,y)$ restricted to the interval $[- La/2, La/2)$. From the second term (it is a part of the mode $m=0$ discarded in \eqref{essa}) arises a logarithmic divergence when $y \rightarrow 0$. To dodge this artefact, we make the choice to remove this 'singular' term and then we instead define the periodized pair-interaction potential energy as:
\begin{equation}
\label{defVcrL}
V_{L,r}(x,y) := \frac{\sqrt{2\pi}}{La} \sum_{m\in \mathbb{Z}^{*}} \mathrm{e}^{i \frac{2\pi}{La} m x} \widehat{V_{r}}(\frac{2\pi}{La}m, y) + \frac{1}{La} \int_{-\frac{La}{2}}^{\frac{La}{2}} \mathrm{d}x'\, V_{r}(x',y),\quad y \neq 0.
\end{equation}
The above pair-interaction potential satisfies the following (we refer the readers to Sec. \ref{PropPoten}):

\begin{lema}
\label{propVcL}
$\forall L \in \mathbb{N}^{*}$ and $\forall 0<2\sqrt{2}r<a$:\\
$\mathrm{(i)}$. $V_{L,r} \in L^{1}(\mathcal{T}_{La,r})$.\\
$\mathrm{(ii)}$. $V_{L,r}(\cdot\,,y)$ is a positive smooth function on $\mathbb{R}$.\\
$\mathrm{(iii)}$. $V_{L,r}(\cdot\,,y)$ converges point-wise to $V_{r}(\cdot\,,y)$ when $L \rightarrow \infty$.
\end{lema}

From the foregoing, introduce the many-body Hamiltonian describing the dynamics of conduction electrons in finite volume systems. For any $L \in \mathbb{N}^{*}$ and $0<2\sqrt{2}r<a$, let $L^{2}(\mathcal{T}_{La,r})$ be the one-particle Hilbert space and $L^{2}(\mathcal{T}_{La,r}^{N}) \cong \otimes_{j=1}^{N} L^{2}(\mathcal{T}_{La,r})$ be the $N$-particle Hilbert space. Note that in our analysis, $N=N_{L}$ and obeys \eqref{relsemic}. We formally consider the family of Hamiltonians:
\begin{equation}
\label{HLr}
\mathcal{H}_{L,r} := \frac{1}{2 m_{e}} \sum_{j=1}^{N} (- \Delta_{x_{j}} -  \Delta_{y_{j}}) + \sum_{j=1}^{N} V_{\mathrm{per}}(x_{j},y_{j}) + \frac{\lambda}{2} \sum_{\substack{j,l = 1 \\ j \neq l}}^{N} V_{L,r}(x_{j}-x_{l},y_{j}-y_{l}),
\end{equation}
where we set $\hbar=1$. $m_{e}$ is the electron rest mass and $\lambda>0$ is a coupling constant. We refer to Sec. \ref{constutt} for a rigorous construction of \eqref{HLr} as a family of self-adjoint operators acting on $L^{2}(\mathcal{T}_{La,r}^{N})$.\\

\indent \textbf{A 1D effective operator.} We continue the modeling by introducing a 1D effective operator to reflect the quasi-1D structure of SWNT. The key idea leading to its derivation is as follows. The operator $\mathcal{H}_{L,r}$ can be represented as a sum of orthogonal transverse modes using the periodic boundary conditions along the circumference of the tube. For small radii of tube $r$, it is reasonable to suppose that the high transverse modes do not contribute much to the low region of the spectrum of \eqref{HLr}. We therefore expect the low-lying spectrum of \eqref{HLr} to be approximated by an effective operator acting on $L^{2}(\mathbb{T}_{aL}^{N})$, obtained from \eqref{HLr} by discarding the high transverse modes. This \textit{ansatz} is based on the works \cite{CDP,CDR1} in which the low-lying spectrum of the Hamiltonian of the 'exciton model' on the infinite-length cylinder $\mathcal{C}_{\infty,r}$ is analyzed. Following the ideas of \cite{CDP,CDR1} and generalizing the method to our actual model, we introduce in \eqref{defHN} below a 1D effective operator which will be the starting-point of our study. Its derivation is outlined in Sec. \ref{ladiscu}. We stress the point that no quantitative analysis justifying that the low-lying spectrum of \eqref{HLr} converges (in a certain sense) to the one of \eqref{defHN} when $r \rightarrow 0$ will be given in this present paper.\\
\indent We introduce some notation. The electric potential energy projected onto the circle is defined by:
\begin{equation}
\label{vEdef}
v_{\mathrm{per}}(x) := \frac{1}{2\pi r} \int_{-\pi r}^{\pi r} \mathrm{d}y\, V_{\mathrm{per}}(x,y) = \frac{1}{2\pi} \int_{-\pi}^{\pi} \mathrm{d}y\, V_{\mathrm{per}}(x,ry),\quad x \in \mathbb{R}.
\end{equation}
Note that $v_{\mathrm{per}}\in L^{2}_{\mathrm{uloc}}(\mathbb{R})$. Since $V_{\mathrm{per}}(x,\cdot\,)$ is $2\pi r$-periodic, then $V_{\mathrm{per}}(x,r\cdot\,)$ is  $2\pi$-periodic. Therefore, $v_{\mathrm{per}}$ is nothing but the average value of the potential energy $V_{\mathrm{per}}$ along the transverse axis. The pair-interaction potential energy in \eqref{Vcrdef} projected onto the circle is defined by:
\begin{equation}
\label{vrdef}
v_{r}(x) := \frac{1}{2\pi r}\int_{-\pi r}^{\pi r} \mathrm{d}y\, V_{r}(x, y)
= \frac{2}{\pi} \frac{e^{2}}{\varepsilon} \frac{1}{\sqrt{x^{2} + 4r^{2}}} \mathpzc{K}(\frac{4r^{2}}{x^{2} + 4 r^{2}}),\quad x \in \mathbb{R}^{*},
\end{equation}
where $\mathpzc{K}$ denotes the complete elliptic integral of the first kind, see \cite{AS}. Note that $v_{r}$ is continuous outside of the origin and admits the following asymptotic expansions:
\begin{gather}
\label{cloO1}
\frac{\varepsilon}{e^{2}} v_{r}(x) = \frac{1}{\pi r}(3\ln(2)+ \ln(r) - \ln(\vert x\vert)) + \frac{1}{16 \pi r^{3}} \vert x\vert^{2} \ln(\vert x\vert) + \mathcal{O}(\vert x\vert^{2})\quad \textrm{when $\vert x\vert \rightarrow 0$},\\
\label{cloO2}
\frac{\varepsilon}{e^{2}} v_{r}(x) = \frac{1}{\vert x\vert} - \frac{r^{2}}{\vert x\vert^{3}} + \mathcal{O}(\frac{1}{\vert x\vert^{4}}) \quad \textrm{when $\vert x\vert \rightarrow \infty$}.
\end{gather}
The periodized Coulomb potential energy in \eqref{defVcrL} projected onto the circle is defined by:
\begin{equation}
\label{vLdef}
\forall L \in \mathbb{N}^{*},\quad v_{L,r}(x) := \frac{1}{2\pi r}\int_{-\pi r}^{\pi r} \mathrm{d}y\, V_{L,r}(x,y),\quad x \in \mathbb{R}\setminus (La \mathbb{Z}).
\end{equation}
$v_{L,r}$ is a symmetric $aL$-periodic function by construction and satisfies the following:

\begin{lema}
\label{propvL}
$\forall L \in \mathbb{N}^{*}$ and $\forall 0<2\sqrt{2}r<a$:\\
$\mathrm{(i)}$. $v_{L,r} \in L^{2}(\mathbb{T}_{La})$ and there exist two constants $c_{1},c_{2}>0$ independent of $L,r$  s.t.
\begin{equation*}
\Vert v_{L,r} \Vert_{2} \leq c_{1}r^{-\frac{1}{2}} + c_{2}.
\end{equation*}
$\mathrm{(ii)}$. $v_{L,r}$ is a positive continuous function on $[-La/2,0[\cup]0,La/2[$.\\
$\mathrm{(iii)}$. The projection onto the circle and the periodization commute: for almost every $x$,
\begin{gather*}
v_{L,r}(x) = \frac{\sqrt{2\pi}}{La} \sum_{m \in \mathbb{Z}^{*}} \mathrm{e}^{i\frac{2\pi}{La} m x} \widehat{v_{r}}(\frac{2\pi}{La}m) + \frac{1}{La} \int_{-\frac{La}{2}}^{\frac{La}{2}} \mathrm{d}x'\, v_{r}(x'),
\end{gather*}
where $\widehat{v_{r}}$ denotes the Fourier transform of $v_{r}$.
\end{lema}

Introduce now the 1D effective Hamiltonian which will be the starting-point of our analysis. Hereafter, the radius of the tube becomes a fixed parameter; say $r= r_{0}>0$ obeying $0<2\sqrt{2} r_{0}<a$ and sufficiently small. We refer to Sec. \ref{ladiscu}. For any $L \in \mathbb{N}^{*}$, let $\mathfrak{h}_{L}:=L^{2}(\mathbb{T}_{La})$ be the one-particle Hilbert space. The $N$-particle Hilbert space is $\mathfrak{h}_{L}^{N}:= L^{2}(\mathbb{T}_{La}^{N}) \cong \otimes_{j=1}^{N} L^{2}(\mathbb{T}_{La})$. In our analysis, $N= N_{L}$ and obeys \eqref{relsemic}. Denoting $\Delta_{L}:= \sum_{j=1}^{N} \mathrm{d}_{x_{j}}^{2}$, define on $\mathcal{C}^{\infty}(\mathbb{T}_{La}^{N})$ the family of operators:
\begin{equation}
\label{defHN}
H_{L} = H_{L,r_{0}} := - \frac{1}{2m_{e}} \Delta_{L} + \sum_{j=1}^{N}  v_{\mathrm{per}}(x_{j}) + \frac{\lambda}{2} \sum_{\substack{j,l =1 \\ j \neq l}}^{N} v_{L,r_{0}}(x_{j} - x_{l}),
\end{equation}
where we set $\hbar=1$. Recall that $\lambda>0$ is a coupling constant. From \eqref{vEdef} along with Lemma \ref{propvL} $\mathrm{(i)}$, the perturbation $H_{L} - (-\frac{1}{2} \Delta_{L})$ is $-\Delta_{L}$-bounded with zero relative bound for any $L \in \mathbb{N}^{*}$. By the Kato-Rellich theorem \cite[Thm. X.19]{RS2}, $H_{L}$ is essentially self-adjoint on $\mathcal{C}^{\infty}(\mathbb{T}_{La}^{N})$ and bounded from below. Its self-adjoint closure, denoted again by $H_{L}$, has domain the Sobolev space: $\mathrm{D}(H_{L}) = \mathcal{W}^{2,2}(\mathbb{T}_{La}^{N})$. Note that by positivity of $v_{L,r_{0}}$, we have the lower bound $H_{L} \geq const(N)$ for some $const(N)\in \mathbb{R}$. Moreover, since the injection $\mathcal{W}^{2,2}(\mathbb{T}_{La}^{N}) \hookrightarrow L^{2}(\mathbb{T}_{La}^{N})$ is compact, then $H_{L}$ has purely discrete spectrum with an accumulation point at infinity.

\subsection{Linear optical response of SWNT--The main result.}

To model the incident light beam in the longitudinal direction of the tube, we consider the following time-dependent electric field:
\begin{equation*}
\mathpzc{E}(t) :=  E \Re \{\mathrm{e}^{i\omega t}\},\quad t \in \mathbb{R},
\end{equation*}
where $E$ denotes the amplitude of the field assumed to be uniform and $\omega>0$ its angular frequency. Without loss of generality, we restrict the amplitude $E$ to the compact interval $[-1,1]$. Introduce an adiabatic switching on of the electric field from time $t_{\mathrm{ini}}=-\infty$ defined as:
\begin{equation*}
\mathcal{E}(t) := \mathrm{e}^{\eta t} \mathpzc{E}(t),\quad t \in \mathbb{R},
\end{equation*}
where $\eta>0$ is the adiabatic parameter. By using Weyl's gauge, the electric field is generated by the time-dependent magnetic vector potential (below, $c$ denotes the speed of light in vacuum):
\begin{gather}
\label{defa2}
A(t):= - c \int_{-\infty}^{t} \mathrm{d}s\, \mathcal{E}(s) = -c E \mathpzc{a}(t),\\
\label{defa}
\mathpzc{a}(t):= \Re (\frac{\mathrm{e}^{(i\omega + \eta) t}}{i\omega + \eta}), \quad t \in \mathbb{R}.
\end{gather}
\indent Introduce the many-body Hamiltonian describing the dynamics of conduction electrons in the presence of the time-dependent electric field. In view of \eqref{defHN} and \eqref{defa}, $\forall L \in \mathbb{N}^{*}$, $\forall E \in [-1,1]$, $\forall \omega>0$ and $\forall \eta>0$, let $\{H_{L}(t),\, t \in \mathbb{R}\}$ be the family of operators on $\mathcal{C}^{\infty}(\mathbb{T}_{La}^{N})$ defined as:
\begin{equation}
\label{HNt}
H_{L}(t) :=  \frac{1}{2 m_{e}}   \sum_{j=1}^{N} (-i \mathrm{d}_{x_{j}} - e E \mathpzc{a}(t))^{2} + \sum_{j=1}^{N} v_{\mathrm{per}}(x_{j}) + \frac{\lambda}{2} \sum_{\substack{j,k =1 \\ j\neq k}}^{N} v_{L,r_{0}}(x_{j} - x_{k}),
\end{equation}
where we set $\hbar=1$. Recall that $N=N_{L}$ obeys \eqref{relsemic}. Denoting $P_{L} := \sum_{j=1}^{N} (- i \mathrm{d}_{x_{j}})$, define:
\begin{equation}
\label{defW}
W_{L}(t) := H_{L}(t) - H_{L} =  - \frac{e}{m_{e}} E \mathpzc{a}(t)P_{L} + \frac{e^{2}}{2m_{e}}N E^{2} \mathpzc{a}^{2}(t),\quad t \in \mathbb{R}.
\end{equation}
$\forall L \in \mathbb{N}^{*}$, $\forall E \in [-1,1]$, $\forall \omega>0$ and $\forall \eta>0$ the perturbation $W_{L}(t)$ is $H_{L}$-bounded with zero relative bound for any $t\in \mathbb{R}$. By the Kato-Rellich Theorem, \eqref{HNt} extends to a family of self-adjoint and bounded from below operators $\forall t \in \mathbb{R}$ with $t$-independent domain $\mathrm{D}(H_{L}(t)) = \mathrm{D}(H_{L})$.\\

From now on, we take into account the fact that the particles obey Fermi-Dirac statistics. Let $P_{-}$ denote the orthogonal projection of $\mathfrak{h}_{L}^{N}:= L^{2}(\mathbb{T}_{La}^{N})$ onto the subspace consisting of antisymmetric functions under permutation of their $N$ arguments. We denote it by $\mathfrak{h}_{L,-}^{N}:=P_{-}\mathfrak{h}_{L}^{N}$. Since the operators $H_{L}$ and $H_{L}(t)$ both commute with $P_{-}$, then they can be restricted to $\mathfrak{h}_{L,-}^{N}$ (they preserve $\mathfrak{h}_{L,-}^{N}$). We denote by $H_{L,-}$ and $H_{L,-}(t)$ respectively the restrictions of $H_{L}$ and $H_{L}(t)$ to $\mathfrak{h}_{L,-}^{N}$. In the same way, we denote by $P_{L,-}$ and $W_{L,-}(t)$ the restrictions of $P_{L}$ and $W_{L}(t)$ to $\mathfrak{h}_{L,-}^{N}$.\\
\indent Next, we turn to the time-evolution of the density matrix associated to the perturbed system.\\
At $t_{\mathrm{ini}}=-\infty$, the system is unperturbed (the electric field is switched off) and assumed to be at thermal equilibrium with a thermal bath. Within the framework of quantum statistical mechanics, the density matrix associated to the unperturbed system is given in the canonical conditions by:
\begin{equation}
\label{densmat-inf}
\rho_{L}^{\mathrm{eq}}(\beta) := (\mathrm{Tr}_{\mathfrak{h}_{L,-}^{N}}\{\mathrm{e}^{-\beta H_{L,-}}\})^{-1} \mathrm{e}^{-\beta H_{L,-}},
\end{equation}
where $\beta := (k_{B} T)^{-1}>0$ is the 'inverse temperature' and $k_{B}$ denotes the Boltzmann constant. Note that \eqref{densmat-inf} is well-defined as trace-class operator on $\mathfrak{h}_{L,-}^{N}$ since the semi-group generated by $H_{L,-}$ is trace-class on $\mathfrak{h}_{L,-}^{N}$, we refer the readers to Sec. \ref{Serrest}. Following Kubo's method, the perturbation by the electric field is adiabatically switched on as the system is brought in time to the present. The time evolution of the density matrix is described by the Liouville-Von Neumann equation. Denoting by $[\cdot\,,\cdot\,]$ the usual commutator, it formally reads as:
\begin{equation}
\label{Liouville}
\left\{\begin{array}{ll}
\displaystyle{i \frac{\partial \rho_{L}}{\partial t}(\beta;t) = [H_{L,-}(t),\rho_{L}(\beta;t)]},\quad t \in \mathbb{R},\\
\lim_{t \rightarrow -\infty} \rho_{L}(\beta;t) = \rho_{L}^{\mathrm{eq}}(\beta).
\end{array}\right.
\end{equation}
Before going further, we need a result of existence and uniqueness of solutions of \eqref{Liouville}. Below, $(\mathfrak{I}_{1}(\mathfrak{h}_{L,-}^{N}),\Vert \cdot\,\Vert_{1})$ denotes the Banach space of trace-class operators on $\mathfrak{h}_{L,-}^{N}$.

\begin{proposition}
\label{soluLoui}
$\forall L \in \mathbb{N}^{*}$, $\forall \beta>0$, $\forall E \in [-1,1]$, $\forall \omega>0$ and $\forall \eta>0$  there exists a family of operators, denoted by $\{\rho_{L}(\beta;t),\, t\in \mathbb{R}\}$, belonging to $\mathfrak{I}_{1}(\mathfrak{h}_{L,-}^{N})$ satisfying the following:\\
$\mathrm{(i)}$. $\rho_{L}(\beta;\cdot\,)$ is differentiable in the trace-norm topology;\\
$\mathrm{(ii)}$. $\rho_{L}(\beta;\cdot\,)$ is the unique solution of the equation \eqref{Liouville} on $\mathfrak{I}_{1}(\mathfrak{h}_{L,-}^{N})$ with initial value $\rho_{L}^{\mathrm{eq}}(\beta)$, i.e.
\begin{equation}
\label{inicondi}
\lim_{t \rightarrow -\infty} \Vert \rho_{L}(\beta;t) - \rho_{L}^{\mathrm{eq}}(\beta)\Vert_{1}=0.
\end{equation}
Moreover, $\rho_{L}(\beta;t)P_{L,-}\in \mathfrak{I}_{1}(\mathfrak{h}_{L,-}^{N})$ for any $t\in \mathbb{R}$.
\end{proposition}

From Proposition \ref{soluLoui}, we can now consider the expectation value of the current operator. We introduce the dependence in $E$ and $\omega,\eta$ in our notation and use $\rho_{L,\eta}(\beta,\omega;E,t)$ instead of $\rho_{L}(\beta;t)$. The statistical quantities that we define below are functions of all those parameters. The current density at a given time $t \in \mathbb{R}$ and at inverse temperature $\beta>0$ induced by the electric field of amplitude $E\in [-1,1]$ and angular frequency $\omega>0$ is defined as, see e.g. \cite{Ma}:
\begin{equation}
\label{current}
\mathcal{J}_{L,\eta}(\beta,\omega;E,t) := - \frac{e}{m_{e} S_{L}} \mathrm{Tr}_{\mathfrak{h}_{L,-}^{N}}\{\rho_{L,\eta}(\beta,\omega;E,t)\sum_{j=1}^{N}(-i\mathrm{d}_{x_{j}} - eE \mathpzc{a}_{\eta}(\omega;t))\},\quad L \in \mathbb{N}^{*},
\end{equation}
where $\mathpzc{a}_{\eta}(\omega;t)=\mathpzc{a}(t)$ is defined in \eqref{defa} and $S_{L}:= 2\pi r_{0} La$. Note that
$\frac{1}{m_{e}}(-i \mathrm{d}_{x} - eE\mathpzc{a}_{\eta}(\omega;t))$ stands for the electron velocity operator. By Proposition \ref{soluLoui}, the trace on the r.h.s. of \eqref{current} is well-defined.\\
\indent Within the linear response theory, the dynamical optical conductivity at $t=0$ and at $\beta>0$ is related to the induced current at $t=0$ by the formal expression, see e.g. \cite[Sec. 3.8]{Ma}:
\begin{equation}
\label{conduc}
\sigma_{L,\eta}(\beta,\omega) := \frac{\partial \mathcal{J}_{L,\eta}}{\partial E}(\beta,\omega;E=0,t=0).
\end{equation}

The main result of this paper is an asymptotic expansion in the low-temperature regime for the dynamical optical conductivity at fixed density of electrons. The leading term that we obtain only involves the eigenvalues and associated eigenfunctions of the unperturbed many-body Hamiltonian $H_{L,-}$. Hereafter, we denote by $\{\mu_{k}\}_{k \geq 0}$, $\mu_{k} = \mu_{k}(L,N)$ (with $N$ obeying \eqref{relsemic}) the set of eigenvalues of $H_{L,-}$ counting multiplicities and in increasing order. We also denote by $\{\psi_{k}\}_{k \geq 0}$ the set of associated normalized eigenvectors. Here is the statement of our main result:

\begin{theorem}
\label{asympA}
$\forall L \in \mathbb{N}^{*}$, $\forall \omega >0$ and $\forall \eta>0$:\\
$\mathrm{(i)}$. The map $E \mapsto \mathcal{J}_{L,\eta}(\beta,\omega;E,t=0)$ is differentiable at $E=0$ for any $\beta>0$.\\
$\mathrm{(ii)}$. There exists a constant $c=c(L,N)>0$ independent of $\beta$ s.t.
\begin{equation}
\label{asymp}
\sigma_{L,\eta}(\beta,\omega) = \varsigma_{L,\eta}(\omega) + \mathcal{O}(\mathrm{e}^{-c \beta}),
\end{equation}
with:
\begin{multline}
\label{defvarsi}
\varsigma_{L,\eta}(\omega) := 4  \frac{e^{2}}{m_{e}^{2} S_{L}} \omega \eta \sum_{k=0}^{\infty} \frac{\vert \langle \psi_{k},P_{L,-} \psi_{0}\rangle \vert^{2}}{[(\mu_{k}-\mu_{0} - \omega)^{2} + \eta^{2}][(\mu_{k} - \mu_{0} + \omega)^{2} + \eta^{2}]} + \\
+ \frac{e^{2}}{m_{e} S_{L}} \frac{\eta}{\omega^{2} + \eta^{2}}\{N - \frac{2}{m_{e}} \sum_{k=0}^{\infty} \frac{(\mu_{k}-\mu_{0}) + 2\omega}{(\mu_{k} - \mu_{0} + \omega)^{2} + \eta^{2}} \vert \langle \psi_{k},P_{L,-} \psi_{0}\rangle \vert^{2}\}.
\end{multline}
\end{theorem}

\begin{remark} The leading term in \eqref{defvarsi} still depends on the adiabatic parameter $\eta$ which is an 'artificial' parameter arising from Kubo's method. From a Physics viewpoint, the relevant quantity is the limit $\eta \downarrow 0$ of \eqref{defvarsi}. In this limit, the second contribution in the r.h.s. of \eqref{defvarsi} identically vanishes. As for the first contribution, it blows up if $\omega = \mu_{k}-\mu_{0}$. This accounts for the sharp peaks observed in the optical absorption spectrum of SWNT, see e.g. \cite{KK,Aj} and references therein. To bring out the presence of peaks, we shall write in the distributional sense:
\begin{equation}
\label{ddrac}
\varsigma_{L}(\omega) := \lim_{\eta \downarrow 0} \varsigma_{L,\eta}(\omega) = \pi \frac{e^{2}}{m_{e}^{2}S_{L}} \frac{1}{\omega} \sum_{k =0}^{\infty} \vert \langle \psi_{k},P_{L,-} \psi_{0}\rangle \vert^{2} \delta(\mu_{k}-\mu_{0} -\omega).
\end{equation}
\end{remark}

\begin{remark} We stress the point that the proof we give of Theorem \ref{asympA} does in no way use the symmetry restriction imposed on the eigenfunctions of the Hamiltonian $H_{L}$. Therefore, the results of Theorem \ref{asympA} still hold true if one considers instead particles obeying Bose-Einstein statistics or Maxwell-Boltzmann statistics (the eigenvalues and eigenfunctions appearing in  \eqref{defvarsi} and \eqref{ddrac} have to be replaced according to the statistics).
\end{remark}

\subsection{Discussion: The 1D effective operator.}
\label{ladiscu}

Here, we outline the derivation of \eqref{defHN} from \eqref{HLr}. As mentioned in Sec. \ref{modl}, for sufficiently small radii of tube $r$, it is reasonable to expect that the high transverse modes do not contribute much to the low region of the spectrum of \eqref{HLr}. This \textit{ansatz} is based on the works \cite{CDP,CDR1} in which the low-lying spectrum of the Hamiltonian of the 'exciton model' on the infinite-length cylinder $\mathcal{C}_{\infty,r}$ is analyzed. On the one hand, it is shown that the low-lying spectrum of the relative motion is only slightly influenced by the high transverses modes for sufficiently small $r$. The low-lying spectrum is then approximated by a 1D effective Hamiltonian and a result of spectrum stability is given in \cite[Sec. 4.2]{CDR1}. On the other hand, it is shown that the low-lying spectrum of this effective Hamiltonian is  well approximated by the spectrum of an operator with point-interactions on the whole line which is exactly solvable. Numerical simulations reinforce this approximation, see \cite{CDP,CDR2}. \\
\indent Following the method used in \cite[Sec. 2.3]{CDR1}, we separate $\mathcal{H}_{L,r}$ into different parts taking advantage of the cylindrical geometry, i.e. we represent $\mathcal{H}_{L,r}$ as a sum of orthogonal transverse modes using the periodic boundary conditions along the circumference of the tube. To do so, consider the orthonormal basis of eigenvectors of $-\frac{1}{2} \Delta_{y}$ with domain $\mathcal{W}^{2,2}(r \mathbb{S})$. By the spectral decomposition:
\begin{equation*}
-\frac{1}{2} \Delta_{y} = \sum_{n \in \mathbb{Z}} \frac{n^{2}}{2r^{2}} \mathpzc{P}_{n}^{r},
\end{equation*}
where the family of 1D orthogonal projections $\{\mathpzc{P}_{n}^{r}\}_{n \in \mathbb{Z}}$ is defined by:
\begin{equation*}
\mathpzc{P}_{n}^{r} := \langle \Phi_{n}^{r},\cdot\,\rangle \Phi_{n}^{r},\quad \Phi_{n}^{r}(y) := \frac{1}{\sqrt{2\pi r}} \mathrm{e}^{i n \frac{y}{r}}.
\end{equation*}
Introduce the family of orthogonal projectors $\Pi_{n}^{r} := \mathbbm{1}\otimes \mathpzc{P}_{n}^{r}$, $n\in \mathbb{Z}$ which project from $L^{2}(\mathcal{T}_{La,r})$ into the $n$-th transverse mode. Turning to the many-body problem, introduce the family $\Pi_{\bold{n}}^{r} := \otimes_{j=1}^{N} \Pi_{n_{j}}^{r}$. Since the projectors are orthogonal, $\mathcal{H}_{L,r}$ can be written as the direct sum:
\begin{equation}
\label{biopl}
\bigoplus_{\bold{n},\bold{m} \in \mathbb{Z}^{N}} \Pi_{\bold{n}}^{r} \mathcal{H}_{L,r} \Pi_{\bold{m}}^{r}.
\end{equation}
The diagonal part ($\bold{n}=\bold{m}$) and off-diagonal part ($\bold{n} \neq \bold{m}$) of the sum read respectively as:
\begin{multline*}
\Pi_{\bold{n}}^{r} \mathcal{H}_{L,r} \Pi_{\bold{n}}^{r} = [(-\frac{1}{2}\Delta_{x_{1}} + v_{\mathrm{per}}(x_{1}) + \frac{n_{1}^{2}}{2 r^{2}})\otimes \mathpzc{P}_{n_{1}}^{r}] \otimes \otimes_{j=2}^{N} \Pi_{n_{j}}^{r} + \dotsb \\
\dotsb + \otimes_{j=1}^{N-1} \Pi_{n_{j}}^{r} \otimes [(-\frac{1}{2} \Delta_{x_{N}} + v_{\mathrm{per}}(x_{N}) + \frac{n_{N}^{2}}{2 r^{2}})\otimes \mathpzc{P}_{n_{N}}^{r}] + \frac{\lambda}{2} \sum_{j\neq l =1}^{N} \Pi_{\bold{n}}^{r} V_{L,r} \Pi_{\bold{n}}^{r},
\end{multline*}
and the last term involves \eqref{vLdef}. If $\bold{n} \neq \bold{m}$, i.e. there exists at least one $j \in \{1,\ldots,N\}$ s.t. $n_{j}\neq m_{j}$:
\begin{equation}
\label{nondago}
\Pi_{\bold{n}}^{r} \mathcal{H}_{L,r} \Pi_{\bold{m}}^{r} = \sum_{j=1}^{N} \Pi_{\bold{n}}^{r} V_{\mathrm{per}} \Pi_{\bold{m}}^{r} +  \frac{\lambda}{2} \sum_{j\neq l =1}^{N} \Pi_{\bold{n}}^{r} V_{L,r} \Pi_{\bold{m}}^{r}.
\end{equation}
We point out that the contributions to the off-diagonal part only come from the potential energies, and they involve terms similar to \eqref{vEdef} and \eqref{vLdef} but with a factor $\mathrm{e}^{i(n_{l} - m_{j})\frac{y}{r}}$ under the integrals. By a natural unitary identification, one can work in the new Hilbert space $\ell^{2}(\mathbb{Z}^{N};L^{2}(\mathbb{T}_{La}^{N}))$ with vectors $\psi = \{\psi_{\bold{n}}\}_{\bold{n} \in \mathbb{Z}^{N}}$, $\psi_{\bold{n}} \in L^{2}(\mathbb{T}_{La}^{N})$. Therefore, the original operator is now an infinite matrix whose elements are operators in $L^{2}(\mathbb{T}_{La}^{N})$. The diagonal matrix elements are given by the operators:
\begin{gather*}
H_{L,r}^{(n_{1},\ldots,n_{N})} := H_{L,r} + \sum_{j=1}^{N} \frac{n_{j}^{2}}{2 r^{2}},\\
H_{L,r} := - \frac{1}{2} \sum_{j=1}^{N} \Delta_{x_{j}} + \sum_{j=1}^{N}  v_{\mathrm{per}}(x_{j}) + \frac{\lambda}{2} \sum_{\substack{j,l =1 \\ j \neq l}}^{N} v_{L,r}(x_{j} - x_{l}).
\end{gather*}
One can see that for $\bold{n} \neq \bold{0}$, the diagonal entries of the infinite operator valued matrix are pushed up by a term proportional with $1/r^{2}$. For $r$ sufficiently small, we thus expect $H_{L,r}$ to be a 'good' candidate for a comparison operator for the low-lying spectrum of $\mathcal{H}_{L,r}$. We mention that an attempt to make this latter statement precise can be found in \cite{ThBJ}. By formally rewriting $\mathcal{H}_{L,r}$ as:
\begin{gather*}
\mathcal{H}_{L,r} = \mathcal{H}_{\mathrm{diag}} + \mathcal{V}_{\mathrm{off-diag}},\\
\mathcal{H}_{\mathrm{diag}} := \bigoplus_{n_{1},\ldots,n_{N} \in \mathbb{Z}} (H_{L,r} + \frac{n_{1}^{2}}{2r^{2}} + \frac{n_{2}^{2}}{2r^{2}} + \dotsb + \frac{n_{N}^{2}}{2r^{2}}),
\end{gather*}
where $\mathcal{V}_{\mathrm{off-diag}}$ contains all the non-diagonal entries (coming from \eqref{nondago}) and zero on the diagonal, we expect $\mathcal{V}_{\mathrm{off-diag}}$ to be relatively form bounded w.r.t. $\mathcal{H}_{\mathrm{diag}}$. Moreover, we expect $\mathcal{V}_{\mathrm{off-diag}}$ to be a 'small' perturbation for sufficiently small $r$. A proof of this statement together with an analysis of the low-lying spectrum of \eqref{defHN} will come in a companion paper.

\section{Proof of Proposition \ref{soluLoui}.}

For the reader's convenience, the proofs of the intermediary results are placed in Sect. \ref{PrIntRe}.\\
Recall that $(\mathfrak{I}_{1}(\mathfrak{h}_{L,-}^{N}),\Vert \cdot\,\Vert_{1})$ denotes the Banach space of trace-class operators on $\mathfrak{h}_{L,-}^{N}$.\\

Dealing with a family of time-dependent Hamiltonians, we need a first result related to the existence of propagators.  We recall that a two-parameter family of unitary operators $\{U(t,s),\, (t,s) \in \mathbb{R}^{2}\}$ is called a unitary propagator if it satisfies the three following conditions, see
\cite[Sec. X.12]{RS2}: $U(r,t)U(t,s) = U(r,s)$, $U(s,s)= \mathbbm{1}$ and $(t,s) \mapsto U(t,s)$ is jointly strongly continuous.\\
Here is a result of existence of propagators associated to the time-dependent Schr\"odinger equation:

\begin{lema}
\label{unitpro}
$\forall L \in \mathbb{N}^{*}$, $\forall E \in [-1,1]$, $\forall \omega>0$ and $\forall \eta>0$ there exists a unitary propagator $U_{L}(t,s)$ on $\mathfrak{h}_{L,-}^{N}$ so that for each $\phi \in \mathrm{D}(H_{L,-})$, $\psi_{s}(t) := U_{L}(t,s) \phi \in \mathrm{D}(H_{L,-})$ for all $t \in \mathbb{R}$ and satisfies:
\begin{equation}
\label{Schrotime}
\left\{\begin{array}{ll}
\displaystyle{i \frac{\mathrm{d} \psi_{s}}{\mathrm{d} t}(t)} &= H_{L,-}(t) \psi_{s}(t) \\
\psi_{s}(s) &= \phi
\end{array}\right..
\end{equation}
In particular, $\mathbb{R} \owns t \mapsto U_{L}(t,s)$ is strongly (continuously) differentiable on $\mathrm{D}(H_{L,-})$ for any $s \in \mathbb{R}$ and is solution of the integral equation taking place  on $\mathrm{D}(H_{L,-})$:
\begin{equation*}
U_{L}(t,s) = \mathbbm{1} - i \int_{s}^{t} \mathrm{d}\tau\, H_{L,-}(\tau) U_{L}(\tau,s),\quad (t,s) \in \mathbb{R}^{2}.
\end{equation*}
\end{lema}

In view of \eqref{defHN} and following Lemma \ref{unitpro}, introduce $\forall L \in \mathbb{N}^{*}$, $\forall E \in [-1,1]$, $\forall \omega>0$ and $\forall \eta>0$ the two-parameter family of unitary operators $\{\Omega_{L}(t,s),\,(t,s) \in \mathbb{R}^{2}\}$ defined on $\mathfrak{h}_{L,-}^{N}$ by:
\begin{equation}
\label{defOmega}
\Omega_{L}(t,s) := \mathrm{e}^{i(t-s) H_{L,-}} U_{L}(t,s),\quad (t,s) \in \mathbb{R}^{2}.
\end{equation}
For each $\phi \in \mathrm{D}(H_{L,-})$, $\Omega_{L}(t,s) \phi \in \mathrm{D}(H_{L,-})$ and $(t,s) \mapsto \Omega_{L}(t,s)$ is jointly strongly continuous.
From \eqref{defW}, introduce also the
family $\{\widetilde{W}_{L,-}(t,s),\, (t,s) \in \mathbb{R}^{2}\}$ defined on $\mathrm{D}(P_{L,-})$ by:
\begin{equation}
\label{tildWN}
\begin{split}
\widetilde{W}_{L,-}(t,s) :&= \mathrm{e}^{i (t-s) H_{L,-}} W_{L,-}(t) \mathrm{e}^{-i(t-s)H_{L,-}} \\
&= - \frac{e}{m_{e}} E \mathpzc{a}(t) \mathrm{e}^{i(t-s)H_{L,-}}P_{L,-} \mathrm{e}^{-i(t-s) H_{L,-}} +  \frac{N}{2 m_{e}} e^{2} E^{2} \mathpzc{a}^{2}(t).
\end{split}
\end{equation}
\indent The second lemma gives some properties on the family of unitary operators defined in
\eqref{defOmega}:

\begin{lema}
\label{propOmega}
$\forall L \in \mathbb{N}^{*}$, $\forall E \in [-1,1]$, $\forall \omega>0$ and $\forall \eta>0$ the operator-valued function $\mathbb{R} \owns t \mapsto \Omega_{L}(t,s)$ is strongly differentiable on $\mathrm{D}(H_{L,-})$ for any $s\in \mathbb{R}$, and its derivative reads as:
\begin{equation*}
\frac{\mathrm{d} \Omega_{L}}{\mathrm{d} t}(t,s) = - i \widetilde{W}_{L,-}(t,s) \Omega_{L}(t,s),\quad (t,s) \in \mathbb{R}^{2}.
\end{equation*}
In particular, $\Omega_{L}(t,s)$ is solution of the integral equation taking place on $\mathrm{D}(H_{L,-})$:
\begin{equation}
\label{relOM}
\Omega_{L}(t,s) = \mathbbm{1} - i \int_{s}^{t} \mathrm{d}\tau\, \widetilde{W}_{L,-}(\tau,s) \Omega_{L}(\tau,s),\quad (t,s) \in \mathbb{R}^{2}.
\end{equation}
\end{lema}

In the third lemma, we investigate the strong limit $t \rightarrow -\infty$ of the family $\{\Omega_{L}(\cdot\,,s),\, s\in \mathbb{R}\}$:

\begin{lema}
\label{stron}
$\forall L \in \mathbb{N}^{*}$, $\forall E \in [-1,1]$, $\forall \omega>0$ and $\forall \eta>0$ the strong limits:
\begin{equation*}
\mathpzc{s}-\lim_{t \rightarrow -\infty} \Omega_{L}(t,s),\quad \mathpzc{s}-\lim_{t \rightarrow -\infty} \Omega^{*}_{L}(t,s),
\end{equation*}
both exist on $\mathfrak{h}_{L,-}^{N}$ for any $s \in \mathbb{R}$. We denote them by $\Omega_{L}^{+}(s)$ and $\Omega_{L}^{*+}(s)$ respectively. Moreover, $\{\Omega_{L}^{+}(s),\, s \in \mathbb{R}\}$ is a one-parameter family of unitary operators satisfying:
\begin{equation}
\label{ulite}
\Omega_{L}^{+*}(s) = \Omega_{L}^{*+}(s),\quad s \in \mathbb{R}.
\end{equation}
\end{lema}


The last lemma ensures in particular that $\Omega_{L}^{+}(s)$, $\Omega_{L}^{*+}(s)$, $s \in \mathbb{R}$ preserve the domain $\mathrm{D}(H_{L,-})$:

\begin{lema}
\label{preservD}
$\forall L \in \mathbb{N}^{*}$, $\forall E \in [-1,1]$, $\forall \omega>0$, $\forall \eta>0$, $\forall \alpha \in \mathbb{R}^{*}$ and $\forall (t,s) \in \mathbb{R}^{2}$ the following four operators are bounded:
\begin{gather*}
(H_{L,-}+i\alpha) \Omega_{L}(t,s) (H_{L,-} + i \alpha)^{-1},\quad (H_{L,-}+i\alpha) \Omega_{L}^{+}(s) (H_{L,-} + i\alpha)^{-1},\\
(H_{L,-}+i\alpha) \Omega_{L}^{*}(t,s) (H_{L,-} + i \alpha)^{-1},\quad (H_{L,-}+i\alpha) \Omega_{L}^{*+}(s) (H_{L,-} + i\alpha)^{-1}.
\end{gather*}
In particular, $\forall L \in \mathbb{N}^{*}$, $\forall \omega>0$, $\forall \eta>0$, $\forall T \geq 0$ and $\forall \alpha \in \mathbb{R}^{*}$ there exists $C>0$ s.t. $\forall E \in [-1,1]$:
\begin{multline}
\label{est11}
\sup_{s,t \in (-\infty,T]} \Vert (H_{L,-}+i\alpha) \Omega_{L}(t,s) (H_{L,-} + i \alpha)^{-1}\Vert + \\
+ \sup_{s,t \in (-\infty,T]} \Vert (H_{L,-}+i\alpha) \Omega_{L}^{*}(t,s) (H_{L,-} + i \alpha)^{-1}\Vert\leq C,
\end{multline}
\begin{equation}
\label{est12}
\sup_{s\in (-\infty,T]} \Vert (H_{L,-}+i\alpha) \Omega_{L}^{+}(s) (H_{L,-} + i \alpha)^{-1}\Vert + \sup_{s\in (-\infty,T]} \Vert (H_{L,-}+i\alpha) \Omega_{L}^{+*}(s) (H_{L,-} + i \alpha)^{-1}\Vert\leq C.
\end{equation}
\end{lema}

We are now ready for the actual proof of Proposition \ref{soluLoui}. From the foregoing, define $\forall L \in \mathbb{N}^{*}$, $\forall \beta>0$, $\forall E \in [-1,1]$, $\forall \omega>0$ and $\forall \eta>0$ the following family of operators:
\begin{equation}
\label{rhot}
\rho_{L}(\beta;t) := \mathrm{e}^{-i t H_{L,-}} \Omega_{L}(t,0) \Omega_{L}^{+*}(0) \rho_{L}^{\mathrm{eq}}(\beta) \Omega_{L}^{+}(0) \Omega_{L}^{*}(t,0) \mathrm{e}^{i t H_{L,-}},\quad t \in \mathbb{R}.
\end{equation}
The rest of this section consists in proving that the family \eqref{rhot} satisfies $\mathrm{(i)}$-$\mathrm{(ii)}$ of Proposition \ref{soluLoui}. Let us first prove that \eqref{rhot} is a family of trace-class operators on $\mathfrak{h}_{L,-}^{N}$. For any $\alpha \in \mathbb{R}^{*}$:
\begin{multline}
\label{rewrhot}
\rho_{L}(\beta;t) = (H_{L,-}+i\alpha)^{-1} \mathrm{e}^{-i t H_{L,-}} \{(H_{L,-}+i\alpha)\Omega_{L}(t,0)(H_{L,-}+i\alpha)^{-1}\} \\
\times \{(H_{L,-}+i\alpha) \Omega_{L}^{+*}(0)(H_{L,-}+i\alpha)^{-1}\} (H_{L,-}+i\alpha)\rho_{L}^{\mathrm{eq}}(\beta)(H_{L,-}+i\alpha) \\
\times \{(H_{L,-}+i\alpha)^{-1} \Omega_{L}^{+}(0)(H_{L,-}+i\alpha)\}\{(H_{L,-}+i\alpha)^{-1}\Omega_{L}^{*}(t,0)(H_{L,-}+i\alpha)\}
\mathrm{e}^{i t H_{L,-}}(H_{L,-}+i\alpha)^{-1}.
\end{multline}
Since the operators between braces are bounded by Lemma \ref{preservD} and $(H_{L,-}+i\alpha)\rho_{L}^{\mathrm{eq}}(\beta)(H_{L,-}+i\alpha)$ is a trace-class operator by Lemma \ref{trass} in Sec. \ref{Serrest}, then \eqref{rhot} is trace-class by the $*$-ideal property of $\mathfrak{I}_{1}(\mathfrak{h}_{L,-}^{N})$. It also follows from \eqref{rewrhot} that $\rho_{L}(\beta;t)P_{L,-} \in \mathfrak{I}_{1}(\mathfrak{h}_{L,-}^{N})$ and  $P_{L,-}\rho_{L}(\beta;t) \in \mathfrak{I}_{1}(\mathfrak{h}_{L,-}^{N})$ by Lemma \ref{relbord} $\mathrm{(i)}$ in Sec. \ref{Serrest}.
Let us secondly prove that \eqref{rhot} satisfies the initial condition \eqref{inicondi}. One has:
\begin{equation*}
\Vert \rho_{L}(\beta;t) - \rho_{L}^{\mathrm{eq}}(\beta)\Vert_{1}
\leq \Vert \Omega_{L}(t,0)\Omega_{L}^{+*}(0)\rho_{L}^{\mathrm{eq}}(\beta)\{\Omega_{L}^{+}(0) \Omega_{L}^{*}(t,0) - \mathbbm{1}\}\Vert_{1} + \Vert \{\Omega_{L}(t,0)\Omega_{L}^{+*}(0) - \mathbbm{1}\} \rho_{L}^{\mathrm{eq}}(\beta)\Vert_{1},
\end{equation*}
where we used that $H_{L,-}$ and $\rho_{L}^{\mathrm{eq}}(\beta)$ commute. Using that:
\begin{gather*}
\Omega_{L}^{+}(0) \Omega_{L}^{*}(t,0) - \mathbbm{1} = \Omega_{L}^{+}(0)(H_{L,-}+i\alpha)\{(H_{L,-}+i\alpha)^{-1}(\Omega_{L}^{*}(t,0) - \Omega_{L}^{+*}(0))\},\\
\Omega_{L}(t,0)\Omega_{L}^{+*}(0) - \mathbbm{1} = \{(\Omega_{L}(t,0) - \Omega_{L}^{+}(0))(H_{L,-}+i\alpha)^{-1}\}(H_{L,-}+i\alpha)\Omega_{L}^{+*}(0),
\end{gather*}
together with \eqref{ulite}, we arrive at:
\begin{multline*}
\Vert \rho_{L}(\beta;t) - \rho_{L}^{\mathrm{eq}}(\beta)\Vert_{1} \leq
\Vert \Omega_{L}(t,0) \Omega_{L}^{+*}(0)\rho_{L}^{\mathrm{eq}}(\beta)(H_{L,-}+i\alpha)\Vert_{1} \times \\
\times \Vert (H_{L,-}+i\alpha)^{-1} \Omega_{L}^{+}(0)(H_{L,-}+i\alpha)\Vert \Vert (H_{L,-}+i\alpha)^{-1}(\Omega_{L}^{*}(t,0) - \Omega_{L}^{*+}(0))\Vert + \\
+ \Vert(\Omega_{L}(t,0) - \Omega_{L}^{+}(0)) (H_{L,-}+i\alpha)^{-1}\Vert \Vert (H_{L,-}+i\alpha)\Omega_{L}^{+*}(0)(H_{L,-}+i\alpha)^{-1}\Vert \Vert (H_{L,-}+i\alpha) \rho_{L}^{\mathrm{eq}}(\beta)\Vert_{1}.
\end{multline*}
On the one hand, $\lim_{t \rightarrow -\infty} \Vert (H_{L,-}+i\alpha)^{-1}(\Omega_{L}^{*}(t,0) - \Omega_{L}^{*+}(0))\Vert = 0 = \lim_{t \rightarrow -\infty} \Vert(\Omega_{L}(t,0) - \Omega_{L}^{+}(0)) (H_{L,-}+i\alpha)^{-1}\Vert$. This follows from Lemma \ref{stron} together with the fact that $(H_{L,-}+i\alpha)^{-1}$ is a compact operator. On the other hand, all the other factors are bounded by Lemmas \ref{preservD} and \ref{trass}. This leads to \eqref{inicondi}.
Let us thirdly prove that \eqref{rhot} is differentiable in the trace-norm topology. Note that $[\rho_{L}(\beta;t),H_{L,-}(t)]$ is bounded, see \eqref{rewrhot} and \eqref{norm1}. Let $t_{0} \in \mathbb{R}$. For $h \in \mathbb{R}^{*}$ small enough:
\begin{equation*}
(\frac{\rho_{L}(\beta;t_{0}+h) - \rho_{L}(\beta;t_{0})}{h} - i [\rho_{L}(\beta;t_{0}),H_{L,-}(t_{0})]) = \mathpzc{Q}_{1}(t_{0},h) + \mathpzc{Q}_{2}(t_{0},h),
\end{equation*}
\begin{multline*}
\mathpzc{Q}_{1}(t_{0},h) := \{\frac{\mathrm{e}^{-i(t_{0}+h) H_{L,-}} - \mathrm{e}^{- i t_{0} H_{L,-}}}{h} \Omega_{L}(t_{0},0)  + \mathrm{e}^{-i (t_{0}+h)H_{L,-}} \frac{\Omega_{L}(t_{0}+h,0) - \Omega_{L}(t_{0},0)}{h} + \\
+ i H_{L,-}(t_{0})\mathrm{e}^{-it_{0} H_{L,-}} \Omega_{L}(t_{0},0)\} \Omega_{L}^{+*}(0) \rho_{L}^{\mathrm{eq}}(\beta) \Omega_{L}^{+}(0) \Omega_{L}^{*}(t_{0},0) \mathrm{e}^{i t_{0} H_{L,-}},
\end{multline*}
\begin{multline*}
\mathpzc{Q}_{2}(t_{0},h) := \mathrm{e}^{-i (t_{0}+h) H_{L,-}} \Omega_{L}(t_{0}+h,0) \Omega_{L}^{+*}(0) \rho_{L}^{\mathrm{eq}}(\beta)  \Omega_{L}^{+}(0) \{\frac{\Omega_{L}^{*}(t_{0}+h,0) - \Omega_{L}^{*}(t_{0},0)}{h} \mathrm{e}^{i t_{0} H_{L,-}} + \\
+ \Omega_{L}^{*}(t_{0}+h,0) \frac{\mathrm{e}^{i(t_{0}+h) H_{L,-}} - \mathrm{e}^{i t_{0} H_{L,-}}}{h} - i \Omega_{L}^{*}(t_{0},0) \mathrm{e}^{i t_{0} H_{L,-}} H_{L,-}(t_{0}) \}+ \\
+ i \{\mathrm{e}^{-i(t_{0}+h) H_{L,-}} \Omega_{L}(t_{0}+h,0) -  \mathrm{e}^{-i t_{0} H_{L,-}} \Omega_{L}(t_{0},0)\} \Omega_{L}^{+*}(0) \rho_{L}^{\mathrm{eq}}(\beta) \Omega_{L}^{+}(0) \Omega_{L}^{*}(t_{0},0) \mathrm{e}^{i t_{0} H_{L,-}}H_{L,-}(t_{0}).
\end{multline*}
We point out that $\forall \alpha \in \mathbb{R}^{*}$, the operator $H_{L,-}(t_{0})\mathrm{e}^{-it_{0} H_{L,-}} \Omega_{L}(t_{0},0) (H_{L,-}+ i\alpha)^{-1}$ is bounded, see Lemma \ref{kyest} along with Lemma \ref{norm2} $\mathrm{(i)}$. Therefore, by Stone's Theorem \cite[Thm. 7.38]{We} and Lemma \ref{propOmega}, it holds on $\mathfrak{h}_{L,-}^{N}$:
\begin{multline*}
\mathpzc{s}-\lim_{h \rightarrow 0} \{\frac{\mathrm{e}^{-i(t_{0}+h) H_{L,-}} - \mathrm{e}^{- i t_{0} H_{L,-}}}{h} \Omega_{L}(t_{0},0) + \mathrm{e}^{-i(t_{0}+h) H_{L,-}} \frac{\Omega_{L}(t_{0}+h,0) - \Omega_{L}(t_{0},0)}{h}\}(H_{L,-}+ i\alpha)^{-1} \\
=- i H_{L,-}(t_{0})\mathrm{e}^{-it_{0} H_{L,-}} \Omega_{L}(t_{0},0) (H_{L,-}+ i\alpha)^{-1}.
\end{multline*}
Since $(H_{L,-} + i\alpha)^{-1}$ is compact, and so is $(H_{L,-}+i\alpha)\Omega_{L}^{+*}(H_{L,-}+i\alpha)^{-2}$ by Lemma \ref{preservD}, then:
\begin{multline}
\label{limnorm}
\lim_{h \rightarrow 0} \Vert \{[\frac{\mathrm{e}^{-i(t_{0}+h) H_{L,-}} - \mathrm{e}^{- i t_{0} H_{L,-}}}{h} \Omega_{L}(t_{0},0) + \mathrm{e}^{- i (t_{0}+h) H_{L,-}} \frac{\Omega_{L}(t_{0}+h,0) - \Omega_{L}(t_{0},0)}{h}] (H_{L,-} + i\alpha)^{-1} \\+ i H_{L,-}(t_{0})\mathrm{e}^{-it_{0} H_{L,-}} \Omega_{L}(t_{0},0) (H_{L,-} + i\alpha)^{-1}\}(H_{L,-}+i\alpha)\Omega_{L}^{+*}(H_{L,-}+i\alpha)^{-2}\Vert = 0.
\end{multline}
We conclude that $\lim_{h \rightarrow 0} \Vert \mathpzc{Q}_{1}(t_{0},h)\Vert_{1} = 0$ since $\Vert (H_{L,-}+i\alpha)^{2} \rho_{L}^{\mathrm{eq}}(\beta) \Omega_{L}^{+}(0) \Omega_{L}^{*}(t_{0},0)\mathrm{e}^{it_{0}H_{L,-}}\Vert_{1} < \infty$, see below \eqref{rewrhot}. We can also prove that $\lim_{h \rightarrow 0}\Vert \mathpzc{Q}_{2}(t_{0},h)\Vert_{1} = 0$ by similar arguments. Therefore:
\begin{equation*}
\lim_{ h \rightarrow 0} \Vert \frac{\rho_{L}(\beta;t_{0}+h) - \rho_{L}(\beta;t_{0})}{h} - i [\rho_{L}(\beta;t_{0}),H_{L,-}(t_{0})]\Vert_{1} = 0.
\end{equation*}
This can be extended to any $t_{0}\in \mathbb{R}$.
To end the proof, it remains to show that \eqref{rhot} is the unique solution of the Von Neumann equation with initial value $\rho_{L}^{\mathrm{eq}}(\beta)$. Assume that there exists another solution, denoted by $\tilde{\rho}_{L}(\beta;\cdot\,)$. Define on $\mathfrak{h}_{L,-}^{N}$ the one-parameter family of unitary operators:
\begin{equation*}
Y_{L}(t) := \mathrm{e}^{- i t H_{L,-}} \Omega_{L}(t,0) \Omega_{L}^{+*}(0),\quad t \in \mathbb{R}.
\end{equation*}
By Lemma \ref{propOmega}, $\mathbb{R} \owns t \mapsto Y_{L}(t)$ is strongly differentiable on $\mathrm{D}(H_{L,-})$ and $\frac{\mathrm{d} Y_{L}}{\mathrm{d}t}(t) = - i H_{L,-}(t) Y_{L}(t)$. For any $\phi, \psi \in \mathrm{D}(H_{L,-})$ and $t \in \mathbb{R}$, one has:
\begin{multline*}
\langle \phi, \frac{\partial}{\partial t} \{Y_{L}^{*}(t)\tilde{\rho}_{L}(\beta;t)Y_{L}(t) - \rho_{L}^{\mathrm{eq}}(\beta)\}\psi\rangle = \\
\langle \phi, i Y_{L}^{*}(t) \{H_{L,-}(t)\tilde{\rho}_{L}(\beta;t) +  [\tilde{\rho}_{L}(\beta;t),H_{L,-}(t)] - \tilde{\rho}_{L}(\beta;t)H_{L,-}(t)\} Y_{L}(t)\psi \rangle = 0,
\end{multline*}
and this can be extended on $\mathfrak{h}_{L,-}^{N}$ by density. Note that $\rho_{L}(\beta;t) = Y_{L}(t)\rho_{L}^{\mathrm{eq}}(\beta)Y_{L}^{*}(t)$. Next use that:
\begin{equation*}
\Vert \tilde{\rho}_{L}(\beta;t) - \rho_{L}(\beta;t)\Vert_{1} \leq \Vert \tilde{\rho}_{L}(\beta;t) - \rho_{L}^{\mathrm{eq}}(\beta)\Vert_{1} + \Vert \rho_{L}^{\mathrm{eq}}(\beta) - \rho_{L}(\beta;t)\Vert_{1},
\end{equation*}
from which we get $\lim_{t \rightarrow - \infty} \Vert \tilde{\rho}_{L}(\beta;t) - \rho_{L}(\beta;t)\Vert_{1}=0$. The uniqueness of the solution is proven.

\section{Proof of Theorem \ref{asympA}.}

For the reader's convenience, the proofs of the intermediary results are placed in Sect. \ref{Sec32}.\\
In this section, we use the notation introduced below Proposition \ref{soluLoui}, see pp. \pageref{current}.\\
\indent We start by the following abstract Lemma:

\begin{lema}
\label{lem30}
$\forall L \in \mathbb{N}^{*}$, $\forall \beta>0$, $\forall E \in [-1,1]$, $\forall \omega>0$, $\forall \eta>0$, $\forall t\in \mathbb{R}$ and for any $k,l \in \{0,1\}$:\\
$\mathrm{(i)}$. $P_{L,-}^{k}\rho_{L}^{\mathrm{eq}}(\beta) P_{L,-}^{l}$ and $[\rho_{L}^{\mathrm{eq}}(\beta),P_{L,-}] \mathrm{e}^{-i t H_{L,-}}P_{L,-}$ are trace-class on $\mathfrak{h}_{L,-}^{N}$.\\
$\mathrm{(ii)}$. $P_{L,-}^{k}\rho_{L,\eta}(\beta,\omega;E,t)P_{L,-}^{l}$ and $[\rho_{L,\eta}(\beta,\omega;E,t),P_{L,-}] \mathrm{e}^{-i t H_{L,-}}P_{L,-}$ are trace-class on $\mathfrak{h}_{L,-}^{N}$.\\
Moreover, their trace norms can be bounded uniformly in $t \in (-\infty,T]$, with $T \geq 0$.
\end{lema}

In view of the formal definition \eqref{conduc}, we need to investigate the behavior in $E$ of \eqref{current}:

\begin{lema}
\label{prop31}
Under the conditions of Lemma \ref{lem30}, the following identity holds on $\mathfrak{I}_{1}(\mathfrak{h}_{L,-}^{N})$:
\begin{equation}
\rho_{L,\eta}(\beta,\omega;E,t)P_{L,-}^{k} = \rho_{L}^{\mathrm{eq}}(\beta) P_{L,-}^{k} + \frac{e}{m_{e}} E \mathcal{R}_{L,\eta}(\beta,\omega;t) P_{L,-}^{k} + \frac{e}{m_{e}} E \mathscr{R}_{L,\eta}(\beta,\omega;E,t)P_{L,-}^{k},
\end{equation}
and each operator on the r.h.s. belongs to $\mathfrak{I}_{1}(\mathfrak{h}_{L,-}^{N})$. $\mathcal{R}_{L,\eta}(\beta,\omega;t)$ is $E$-independent and reads as:
\begin{equation}
\label{R1}
\mathcal{R}_{L,\eta}(\beta,\omega;t) := - i \int_{-\infty}^{t} \mathrm{d}\tau\, \mathpzc{a}_{\eta}(\omega;\tau) \mathrm{e}^{i(\tau-t)H_{L,-}}[\rho_{L}^{\mathrm{eq}}(\beta),P_{L,-}] \mathrm{e}^{-i(\tau-t)H_{L,-}},
\end{equation}
where $\mathpzc{a}_{\eta}(\omega;t)=\mathpzc{a}(t)$ is given in \eqref{defa}, as for the last term, it is defined as:
\begin{equation}
\label{R2}
\mathscr{R}_{L,\eta}(\beta,\omega;E,t) := - i \int_{-\infty}^{t} \mathrm{d}\tau\, \mathpzc{a}_{\eta}(\omega;\tau) \mathrm{e}^{i(\tau-t)H_{L,-}}[(\rho_{L,\eta}(\beta,\omega;E,\tau) - \rho_{L}^{\mathrm{eq}}(\beta)),P_{L,-}] \mathrm{e}^{-i(\tau-t)H_{L,-}}.
\end{equation}
\end{lema}

From \eqref{current}, one infers from Lemma \ref{prop31} the following formula for the induced current at $t=0$:
\begin{multline}
\label{current0}
m_{e}^{2}S_{L} \mathcal{J}_{L,\eta}(\beta,\omega;E,0) =
- e m_{e}  \mathrm{Tr}_{\mathfrak{h}_{L,-}^{N}}\{\rho_{L}^{\mathrm{eq}}(\beta) P_{L,-}\}  \\
 + e^{2} E [m_{e} N \mathpzc{a}_{\eta}(\omega;0)  - \mathrm{Tr}_{\mathfrak{h}_{L,-}^{N}}\{\mathcal{R}_{L,\eta}(\beta,\omega;0)P_{L,-}\}]
- e^{2}E \mathrm{Tr}_{\mathfrak{h}_{L,-}^{N}}\{\mathscr{R}_{L,\eta}(\beta,\omega;E,0)P_{L,-}\} \\ + e^{3}E^{2} N \mathpzc{a}_{\eta}(\omega;0)[\mathrm{Tr}_{\mathfrak{h}_{L,-}^{N}}\{\mathcal{R}_{L,\eta}(\beta,\omega;0)\}+ \mathrm{Tr}_{\mathfrak{h}_{L,-}^{N}}\{\mathscr{R}_{L,\eta}(\beta,\omega;E,0)\}].
\end{multline}
Since the operator $\mathscr{R}_{L,\eta}(\beta,\omega;E,0)$ can also be rewritten as (see \eqref{rhoLnew}):
\begin{multline*}
-\frac{e}{m_{e}} E \int_{-\infty}^{0} \mathrm{d}\tau \int_{-\infty}^{\tau} \mathrm{d}s\, \mathpzc{a}_{\eta}(\omega;\tau) \mathpzc{a}_{\eta}(\omega;s) \times \\
\times  \mathrm{e}^{i\tau H_{L,-}} [\mathrm{e}^{i(s-\tau)H_{L,-}}[\rho_{L}(\beta,\omega;E,s),P_{L,-}] \mathrm{e}^{-i(s-\tau)H_{L,-}}, P_{L,-}] \mathrm{e}^{-i\tau H_{L,-}},
\end{multline*}
which holds in the trace-class operators sense by Lemma \ref{lem30} $\mathrm{(ii)}$, then we expect the third and fourth contribution in the r.h.s. of \eqref{current0} to behave like $\mathcal{O}(E^{2})$ when $E \rightarrow 0$. Indeed:

\begin{lema}
\label{prop32}
$\forall L \in \mathbb{N}^{*}$, $\forall \beta>0$, $\forall \omega>0$, $\forall \eta>0$ and for any $k \in \{0,1\}$  there exists $C>0$ s.t.:
\begin{equation}
\label{labon}
\forall E \in [-1,1],\quad \vert  \mathrm{Tr}_{\mathfrak{h}_{L,-}^{N}}\{\mathscr{R}_{L,\eta}(\beta,\omega;E,0) P_{L,-}^{k}\} \vert \leq C \vert E \vert (1+\vert E\vert).
\end{equation}
Moreover, the map $E \mapsto E \mathrm{Tr}_{\mathfrak{h}_{L,-}^{N}}\{\mathscr{R}_{L,\eta}(\beta,\omega;E,0) P_{L,-}^{k}\}$ is differentiable at $E=0$, and one has:
\begin{equation*}
\frac{\partial}{\partial E} (E \times \mathrm{Tr}_{\mathfrak{h}_{L,-}^{N}}\{\mathscr{R}_{L,\eta}(\beta,\omega;E,0)P_{L,-}^{k}\})\Big\vert_{E=0} = 0.
\end{equation*}
\end{lema}

We are now ready for the actual proof of Theorem \ref{asympA}. $\mathrm{(i)}$ directly follows from \eqref{current0} along with Lemma \ref{prop32}. We turn to $\mathrm{(ii)}$. In view of \eqref{conduc}, we obtain from \eqref{current0} along with Lemma \ref{prop32}:
\begin{equation}
\label{absorbv}
\sigma_{L,\eta}(\beta,\omega) = - \frac{e^{2}}{m_{e}^{2}S_{L}} \mathrm{Tr}_{\mathfrak{h}_{L,-}^{N}}\{\mathcal{R}_{L,\eta}(\beta,\omega;0)P_{L,-}\} + \frac{e^{2}}{m_{e}}\frac{N}{S_{L}}\mathpzc{a}_{\eta}(\omega;0).
\end{equation}
Let us rewrite the  first term on the r.h.s. of \eqref{absorbv} in terms of eigenvalues $\{\mu_{k}\}_{k \geq 0}$, $\mu_{k} = \mu_{k}(L,N)$ (counting multiplicities and in increasing order)  and associated normalized eigenvectors $\{\psi_{k}\}_{k \geq 0}$ of $H_{L,-}$. From \eqref{R1}, the quantity $\mathrm{Tr}_{\mathfrak{h}_{L,-}^{N}}\{\mathcal{R}_{L,\eta}(\beta,\omega;0) P_{L,-}\}$ can be rewritten as:
\begin{equation*}
i \int_{-\infty}^{0} \mathrm{d}\tau\, \mathpzc{a}_{\eta}(\omega;\tau) \mathrm{Tr}_{\mathfrak{h}_{L,-}^{N}}\{P_{L,-} \mathrm{e}^{i \tau H_{L,-}} P_{L,-} \mathrm{e}^{-i \tau H_{L,-}} \rho_{L}^{\mathrm{eq}}(\beta)- \rho_{L}^{\mathrm{eq}}(\beta) \mathrm{e}^{i \tau H_{L,-}} P_{L,-} \mathrm{e}^{-i \tau H_{L,-}}P_{L,-}\},
\end{equation*}
where we used that $[\rho_{L}^{\mathrm{eq}}(\beta),H_{L,-}]=0$ together with the cyclicity property of the trace. Since the $\psi_{k}$s form an orthonormal basis in $\mathfrak{h}_{L,-}^{N}$, then one obtains from the spectral theorem:
\begin{multline*}
\mathrm{Tr}_{\mathfrak{h}_{L,-}^{N}}\{\mathcal{R}_{L,\eta}(\beta,\omega;0)P_{L,-}\}
= \\ -\frac{2}{Z_{L}(\beta)} \int_{-\infty}^{0} \mathrm{d}\tau\, \mathpzc{a}_{\eta}(\omega;\tau) \sum_{k=0}^{\infty} \mathrm{e}^{-\beta \mu_{k}} \Im \{\mathrm{e}^{-i\mu_{k} \tau} \langle P_{L,-} \psi_{k}, \mathrm{e}^{i \tau H_{L,-}} P_{L,-} \psi_{k} \rangle\},
\end{multline*}
where $Z_{L}(\beta)$ denotes the canonical partition function. Define the reduced partition function as:
\begin{equation}
\label{partmod}
\forall \beta>0,\quad \tilde{Z}_{L}(\beta) := \mathrm{e}^{\beta \mu_{0}}Z_{L}(\beta) = \mathrm{e}^{\beta \mu_{0}} \mathrm{Tr}_{\mathfrak{h}_{L,-}^{N}}\{\mathrm{e}^{-\beta H_{L,-}}\}.
\end{equation}
By involving \eqref{partmod}, we arrive at the following rewriting:
\begin{gather}
\mathrm{Tr}_{\mathfrak{h}_{L,-}^{N}}\{\mathcal{R}_{L,\eta}(\beta,\omega;0) P_{L,-}\} = \mathpzc{T}_{L,\eta}^{(0)}(\beta,\omega) + \mathpzc{T}_{L,\eta}^{(1)}(\beta,\omega), \nonumber \\
\label{T0}
\mathpzc{T}_{L,\eta}^{(0)}(\beta,\omega) := -\frac{2}{\tilde{Z}_{L}(\beta)} \int_{-\infty}^{0} \mathrm{d}\tau\, \mathpzc{a}_{\eta}(\omega;\tau) \Im \{\mathrm{e}^{-i\mu_{0} \tau} \langle P_{L,-} \psi_{0}, \mathrm{e}^{i \tau H_{L,-}} P_{L,-} \psi_{0} \rangle\}, \\
\label{T1}
\mathpzc{T}^{(1)}_{L,\eta}(\beta,\omega) := -\frac{2}{\tilde{Z}_{L}(\beta)} \int_{-\infty}^{0} \mathrm{d}\tau\, \mathpzc{a}_{\eta}(\omega;\tau) \sum_{k=1}^{\infty} \mathrm{e}^{-\beta(\mu_{k} - \mu_{0})} \Im \{\mathrm{e}^{-i \mu_{k}\tau} \langle P_{L,-} \psi_{k}, \mathrm{e}^{i \tau H_{L,-}} P_{L,-} \psi_{k} \rangle\}.
\end{gather}
The rest of the proof consists in showing that \eqref{T0} is the leading term of the expansion in \eqref{asymp}, and \eqref{T1} is exponentially decreasing in $\beta$. By the spectral decomposition of $\mathrm{e}^{i \tau H_{L,-}}$, one has:
\begin{equation*}
\mathpzc{T}^{(0)}_{L,\eta}(\beta,\omega) = -\frac{2}{\tilde{Z}_{L}(\beta)} \sum_{k=0}^{\infty} \vert \langle \psi_{k}, P_{L,-} \psi_{0} \rangle\vert^{2} \int_{-\infty}^{0} \mathrm{d}\tau\, \mathpzc{a}_{\eta}(\omega;\tau) \Im\{\mathrm{e}^{i(\mu_{k} - \mu_{0}) \tau}\},
\end{equation*}
and straightforward calculations lead to:
\begin{equation*}
\mathpzc{T}^{(0)}_{L,\eta}(\beta,\omega) = \frac{2 \eta}{\omega^{2} + \eta^{2}} \frac{1}{\tilde{Z}_{L}(\beta)}  \sum_{k=0}^{\infty} \frac{(\mu_{k} - \mu_{0})\{(\mu_{k}-\mu_{0})^{2}+ \eta^{2} - 3 \omega^{2}\} }{[(\mu_{k} - \mu_{0} - \omega)^{2} + \eta^{2}][(\mu_{k} - \mu_{0} + \omega)^{2} + \eta^{2}]} \vert\langle \psi_{k},P_{L,-} \psi_{0}\rangle\vert^{2}.
\end{equation*}
It remains to use the identities:
\begin{gather*}
(\mu_{k}-\mu_{0})^{2} + \eta^{2} - 3 \omega^{2} = [(\mu_{k} - \mu_{0} - \omega)^{2} + \eta^{2}] + 2 \omega (\mu_{k} - \mu_{0}) - 4 \omega^{2};\\
(\mu_{k}-\mu_{0})\{2 \omega (\mu_{k} - \mu_{0}) - 4 \omega^{2}\} = 2\omega[(\mu_{k} - \mu_{0} - \omega)^{2} + \eta^{2}] - 2\omega(\eta^{2} + \omega^{2}).
\end{gather*}
As for the quantity in \eqref{T1}, we have the following estimate concluding the proof of Theorem \ref{asympA}:

\begin{lema}
\label{lem31}
$\forall L \in \mathbb{N}^{*}$, $\forall \omega>0$ and $\forall \eta>0$ there exist three constants $C_{l}=C_{l}(L,N,\omega,\eta)>0$, $l=1,2$  and $c=c(L,N)>0$ s.t. for $\beta$ sufficiently large:
\begin{equation*}
\vert \mathpzc{T}^{(1)}_{L,\eta}(\beta,\omega)\vert \leq C_{1} \mathrm{e}^{-\beta(\mu_{1}-\mu_{0})} + C_{2} \frac{\mathrm{e}^{-c \beta}}{\beta}.
\end{equation*}
\end{lema}

\section{Proof of intermediary results.}

\subsection{Proof of Lemma \ref{unitpro}--\ref{preservD}.}
\label{PrIntRe}

For simplicity's sake, we set $m_{e}=1$ and $e=1$ in the definitions \eqref{defW} and \eqref{tildWN}.\\

\noindent \textbf{Proof of Lemma \ref{unitpro}.} Let $L \in \mathbb{N}^{*}$, $E \in [-1,1]$, $\omega>0$ and $\eta>0$ be fixed. It is enough to verify the assumptions of \cite[Thm. 9.5.3]{BEH}, see also \cite[Thm. X.70]{RS2}.
$\mathrm{(i)}$. The family $\{W_{L,-}(t),\, t \in \mathbb{R}\}$ is $H_{L,-}$-bounded with zero relative bound.
$\mathrm{(ii)}$. By Lemma \ref{estimnorm1}, define in the bounded operators sense:
\begin{equation*}
\mathpzc{C}(t,s) := (H_{L,-}(s) - i)(H_{L,-}(t) + i)^{-1} - \mathbbm{1} = \{W_{L,-}(t) - W_{L,-}(s)\}(H_{L,-}(s)-i)^{-1},\quad (s,t) \in \mathbb{R}^{2}.
\end{equation*}
Let us note first that, in view of \eqref{defW}, the family $\{W_{L,-}(t),\, t \in \mathbb{R}\}$ is strongly continuously differentiable on $\mathrm{D}(P_{L,-})$. Define then on $\mathrm{D}(P_{L,-})$ the following family of operators:
\begin{equation}
\label{derWN}
W_{L,-}'(t) := - E \mathpzc{a}'(t) P_{L,-} + N E^{2} \mathpzc{a}'(t) \mathpzc{a}(t),\quad t \in \mathbb{R}.
\end{equation}
It follows in particular that $t \mapsto W_{L,-}(t) (H_{L,-}(t_{0}) - i)^{-1}$ is continuous in norm topology. Secondly, it is easy to see from \eqref{defa} that $\forall K \subset \mathbb{R}$ compact subset there exists $C>0$ s.t. $\forall s,t \in K$:
\begin{equation*}
\vert \mathpzc{a}(s) - \mathpzc{a}(t) \vert + \vert \mathpzc{a}^{2}(s) - \mathpzc{a}^{2}(t)\vert \leq C \vert s-t\vert.
\end{equation*}
For any $(t_{j},s_{j}) \in \mathbb{R}^{2}$, with $t_{j} \neq s_{j}$, $j=0,1$ introduce:
\begin{multline*}
\frac{\mathpzc{C}(t_{1},s_{1})}{t_{1}-s_{1}} - \frac{\mathpzc{C}(t_{0},s_{0})}{t_{0}-s_{0}} =
\{\frac{W_{L,-}(t_{1})}{t_{1}-s_{1}} - \frac{W_{L,-}(t_{0})}{t_{0}-s_{0}}\}(H_{L,-}(s_{0})-i)^{-1} + \\
- \{\frac{W_{L,-}(s_{1})}{t_{1}-s_{1}} - \frac{W_{L,-}(s_{0})}{t_{0}-s_{0}}\}(H_{L,-}(s_{0})-i)^{-1} +\\
- \frac{W_{L,-}(t_{1}) - W_{L,-}(s_{1})}{t_{1}-s_{1}}(H_{L,-}(s_{1}) - i)^{-1} \{W_{L,-}(s_{1}) - W_{L,-}(s_{0})\}(H_{L,-}(s_{0})-i)^{-1}.
\end{multline*}
By standard arguments and the properties mentioned above along with \eqref{norm1}-\eqref{norm2}, we obtain:
$$\lim_{(t_{1},s_{1}) \rightarrow (t_{0},s_{0})} \Vert \frac{\mathpzc{C}(t_{1},s_{1})}{t_{1}-s_{1}} - \frac{\mathpzc{C}(t_{0},s_{0})}{t_{0}-s_{0}}\Vert = 0.$$ Thus, $(t,s) \mapsto (t-s)^{-1} \mathpzc{C}(t,s)$ is strongly uniformly continuous on $\mathfrak{h}_{L,-}^{N}$ if $s,t$ belong to a compact subset and $s\neq t$.
$\mathrm{(iii)}$. For any $\phi \in \mathfrak{h}_{L,-}^{N}$, $\mathpzc{C}(t) \phi := \lim_{s \rightarrow t} (t-s)^{-1} \mathpzc{C}(t,s) \phi$ exists uniformly in each compact subset of $\mathbb{R}$, and
$t \mapsto \mathpzc{C}(t) = W_{L,-}'(t) (H_{L,-}(t) - i)^{-1}$ is continuous in the norm topology. \qed \\

\noindent \textbf{Proof of Lemma \ref{propOmega}.} Let $s \in \mathbb{R}$. For any $t_{0} \in \mathbb{R}$, $h \in \mathbb{R}^{*}$ small enough and $\varphi \in \mathrm{D}(H_{L,-})$:
\begin{equation*}
\begin{split}
\frac{\Omega_{L}(t_{0}+ h,s) - \Omega_{L}(t_{0},s)}{h} \varphi &= \frac{\mathrm{e}^{i(t_{0}+h-s) H_{L,-}} - \mathrm{e}^{i(t_{0} -s)H_{L,-}}}{h} U_{L}(t_{0}+h,s) \varphi \\
&+ \mathrm{e}^{i(t_{0}-s)H_{L,-}} \frac{U_{L}(t_{0}+h,s) - U_{L}(t_{0},s)}{h} \varphi.
\end{split}
\end{equation*}
By using Stone's theorem \cite[Thm. 7.38]{We} for the first term, and the fact that $\mathbb{R} \owns t \mapsto U_{L}(t,s)$ is strongly differentiable on $\mathrm{D}(H_{L,-})$ for the second term along with \eqref{Schrotime}:
\begin{align*}
\lim_{h \rightarrow 0} \frac{\Omega_{L}(t_{0}+ h,s) - \Omega_{L}(t_{0},s)}{h} \varphi &= i \mathrm{e}^{i(t_{0}-s)H_{L,-}} H_{L,-} U_{L}(t_{0},s) \varphi + \mathrm{e}^{i(t_{0} -s) H_{L,-}} \frac{\mathrm{d} U_{L}}{\mathrm{d} t}(t_{0},s) \varphi \\
&= i \mathrm{e}^{i(t_{0}-s)H_{L,-}} (H_{L,-} - H_{L,-}(t_{0}))\mathrm{e}^{-i(t_{0}-s)H_{L,-}} \mathrm{e}^{i(t_{0}-s)H_{L,-}} U_{L}(t_{0},s) \varphi\\
&= -i \widetilde{W}_{L,-}(t_{0},s) \Omega_{L}(t_{0},s) \varphi,
\end{align*}
where we used \eqref{tildWN}. Since $\mathbb{R} \owns t \mapsto \widetilde{W}_{L,-}(t,s)\Omega_{L}(t,s)$ is strongly continuous on $\mathrm{D}(H_{L,-})$, we get:
\begin{equation*}
\Omega_{L}(t,s) = \Omega_{L}(s,s) - i \int_{s}^{t} \mathrm{d}\tau\, \widetilde{W}_{L,-}(\tau,s) \Omega_{L}(\tau,s)  = \mathbbm{1} - i \int_{s}^{t}  \mathrm{d}\tau\, \widetilde{W}_{L,-}(\tau,s) \Omega_{L}(\tau,s). \tag*{\qed}
\end{equation*}

Before turning to the proof of Lemmas \ref{stron}-\ref{preservD}, we need the following:

\begin{lema}
\label{kyest}
$\forall L \in \mathbb{N}^{*}$, $\forall E \in [-1,1]$, $\forall \omega>0$, $\forall \eta>0$ and $\forall \alpha \in \mathbb{R}^{*}$ the following two-parameter family of operators is bounded:
\begin{equation*}
(H_{L,-}(t) + i \alpha) U_{L}(t,s) (H_{L,-}(t) + i\alpha)^{-1},\quad (t,s) \in \mathbb{R}^{2}.
\end{equation*}
In particular, $\forall L \in \mathbb{N}^{*}$, $\forall \omega>0$, $\forall \eta>0$, $\forall T \geq 0$ and $\forall \alpha \in \mathbb{R}^{*}$ there exists $C>0$ s.t. $\forall E \in [-1,1]$:
\begin{equation}
\label{unifet}
\sup_{t,s \in (-\infty,T]} \Vert (H_{L,-}(t) + i \alpha) U_{L}(t,s) (H_{L,-}(t) + i\alpha)^{-1} \Vert \leq C.
\end{equation}
\end{lema}

\noindent \textbf{Proof of Lemma \ref{kyest}.} Introduce the following two-parameter family of operators:
\begin{equation*}
\mathpzc{T}_{s}(t) := U_{L}^{*}(t,s) (H_{L,-}(t) + i\alpha)U_{L}(t,s) (H_{L,-}(t) + i\alpha)^{-1},\quad (t,s) \in \mathbb{R}^{2}.
\end{equation*}
Since $\forall \varphi \in \mathrm{D}(H_{L,-})$, $U_{L}(t,s) \varphi \in \mathrm{D}(H_{L,-})$ and $H_{L,-}(t)$ is closed on $\mathrm{D}(H_{L,-})$, then $\mathpzc{T}_{s}(t)$ is bounded by the closed graph theorem. 
The difficulty consists in deriving an upper bound uniform in $t,s \in (-\infty,T]$.
For any $\phi \in \mathrm{D}(H_{L,-})$ and $\psi \in \mathfrak{h}_{L,-}^{N}$ both with norm one, and $s \in \mathbb{R}$, introduce on $\mathbb{R}$ the function:
\begin{equation*}
\mathpzc{f}_{s}(t) := \langle \phi, \mathpzc{T}_{s}(t) \psi \rangle.
\end{equation*}
Let $t_{0}\in \mathbb{R}$ and pick $h\in \mathbb{R}^{*}$ small enough. For any $\alpha \in \mathbb{R}^{*}$:
\begin{multline}
\label{whol}
\frac{\mathpzc{f}_{s}(t_{0}+h) - \mathpzc{f}_{s}(t_{0})}{h} \\
\begin{split}
&= \langle \frac{U_{L}(t_{0}+h,s) - U_{L}(t_{0},s)}{h} \phi, (H_{L,-}(t_{0}+h) + i\alpha)U_{L}(t_{0}+h,s)(H_{L,-}(t_{0}+h) + i\alpha)^{-1}\psi \rangle \\
&+ \langle U_{L}(t_{0},s) \phi, \frac{H_{L,-}(t_{0}+h) - H_{L,-}(t_{0})}{h}U_{L}(t_{0}+h,s)(H_{L,-}(t_{0}+h) + i\alpha)^{-1}\psi \rangle \\
&+ \langle (H_{L,-}(t_{0}) + i\alpha) U_{L}(t_{0},s) \phi, \frac{U_{L}(t_{0}+h,s) - U_{L}(t_{0},s)}{h} (H_{L,-}(t_{0}+h) + i\alpha)^{-1}\psi \rangle \\
&+ \langle U_{L}(t_{0},s)\phi, (H_{L,-}(t_{0}) + i\alpha)U_{L}(t_{0},s)\frac{(H_{L,-}(t_{0}+h) + i\alpha)^{-1} - (H_{L,-}(t_{0}) + i\alpha)^{-1}}{h}\psi \rangle.
\end{split}
\end{multline}
Since $\forall \varphi \in \mathrm{D}(H_{L,-})$, $\lim_{h \rightarrow 0} h^{-1}(U_{L}(t_{0}+h,s) - U_{L}(t_{0},s))\varphi = -i H_{L,-}(t_{0})U_{L}(t_{0},s) \varphi$ by Lemma \ref{unitpro}, then by taking the limit $h \rightarrow 0$ in \eqref{whol} the first and third terms cancel out each other. From \eqref{derWN}
and by introducing the family of operators $\{G_{s}(t),\, (t,s) \in \mathbb{R}^{2}\}$ defined as:
\begin{equation}
\label{Gst}
G_{s}(t) := U_{L}^{*}(t,s) W_{L,-}'(t) (H_{L,-}(t) + i\alpha)^{-1} U_{L}(t,s) \mathpzc{T}_{s}(t) - \mathpzc{T}_{s}(t) W_{L,-}'(t) (H_{L,-}(t) + i\alpha)^{-1},
\end{equation}
we are then left with:
\begin{equation*}
\mathpzc{f}_{s}'(t_{0}):= \lim_{h \rightarrow 0} \frac{\mathpzc{f}_{s}(t_{0}+h) - \mathpzc{f}_{s}(t_{0})}{h} = \langle \phi, G_{s}(t_{0}) \psi\rangle.
\end{equation*}
Such a result can be extended to any $t_{0} \in \mathbb{R}$, and $\mathpzc{f}_{s}$ is therefore differentiable on $\mathbb{R}$. Note that the family $G_{s}(t)$ defined in \eqref{Gst} is bounded on $\mathfrak{h}_{L,-}^{N}$ by Lemmas \ref{relbord}-\ref{estimnorm1} and the foregoing.
By using that $\mathbb{R} \owns t \mapsto G_{s}(t)$ is strongly continuous on $\mathfrak{h}_{L,-}^{N}$, one has:
\begin{equation*}
\mathpzc{f}_{s}(t) = \mathpzc{f}_{s}(s) + \int_{s}^{t} \mathrm{d}\tau\, \mathpzc{f}_{s}'(\tau).
\end{equation*}
Now restrict to $s,t \in (-\infty,T]$ and suppose (without loss of generality) that $s<t\leq T$. We have:
\begin{equation*}
\vert \mathpzc{f}_{s}(t)\vert \leq 1 + \int_{s}^{T} \mathrm{d}\tau\, \Vert G_{s}(\tau)\Vert.
\end{equation*}
By a density argument, it follows that:
\begin{equation*}
\Vert \mathpzc{T}_{s}(t)\Vert \leq 1 +  \int_{s}^{T}\mathrm{d}\tau\, \Vert G_{s}(\tau)\Vert.
\end{equation*}
From \eqref{derWN} and by using Lemmas \ref{relbord}-\ref{estimnorm1}, there exists $C>0$ s.t. $\forall s,\tau \in (-\infty,T]$ and $\forall E \in [-1,1]$:
\begin{equation*}
\Vert G_{s}(\tau) \Vert \leq C \vert \mathpzc{a}'(\tau)\vert (1 + N \vert \mathpzc{a}(\tau)\vert)\Vert \mathpzc{T}_{s}(\tau)\Vert.
\end{equation*}
Setting $\mathpzc{g}(u):= C \vert \mathpzc{a}'(u)\vert (1 + N \vert \mathpzc{a}(u)\vert)>0$, Gr\"onwall's inequality enables us to get:
\begin{equation*}
\Vert \mathpzc{T}_{s}(t)\Vert \leq \exp(\int_{s}^{T} \mathrm{d}u\, \mathpzc{g}(u)).
\end{equation*}
It remains to extend the integration over $u$ on $(-\infty,T]$ and the proof is over.   \qed \\

\noindent \textbf{Proof of Lemma \ref{stron}.} We start the proof by the following estimate. By Lemmas \ref{relbord}, \ref{estimnorm1} and \ref{kyest}, for any $s \in \mathbb{R}$, $\alpha \in \mathbb{R}^{*}$ and $t \in (-\infty,T]$, with $T\geq 0$:
\begin{multline*}
\Vert P_{L,-} U_{L}(t,s)(H_{L,-}+ i\alpha)^{-1}\Vert =
\Vert P_{L,-} (H_{L,-}+ i\alpha)^{-1} \Vert \Vert (H_{L,-}+ i\alpha) (H_{L,-}(t) + i\alpha)^{-1}\Vert \\ \times \Vert (H_{L,-}(t) + i\alpha) U_{L}(t,s) (H_{L,-}(t) + i\alpha)^{-1} \Vert \Vert (H_{L,-}(t) + i\alpha) (H_{L,-}+ i\alpha)^{-1}\Vert.
\end{multline*}
By using \eqref{normPjres}, \eqref{norm2} along with \eqref{unifet}, there exists $C>0$ s.t. $\forall s \in \mathbb{R}$ and $\forall E \in [-1,1]$:
\begin{equation}
\label{usefetl}
\sup_{t \in (-\infty,T]}  \Vert P_{L,-} U_{L}(t,s)(H_{L,-}+ i\alpha)^{-1}\Vert \leq C.
\end{equation}
Let us now turn to the actual proof. Let $t_{1}, t_{2} \in \mathbb{R}$ s.t. $t_{1}<t_{2}$. Let $T\geq 0$ s.t. $t_{2} < T$. From \eqref{relOM}, one has for any $s \in \mathbb{R}$, $E \in [-1,1]$ and $\varphi \in \mathrm{D}(H_{L,-})$:
\begin{equation*}
(\Omega_{L}(t_{1},s) - \Omega_{L}(t_{2},s)) \varphi = \int_{t_{1}}^{t_{2}} \mathrm{d}\tau\, \widetilde{W}_{L,-}(\tau,s) \Omega_{L}(\tau,s) \varphi.
\end{equation*}
Let $\alpha \in \mathbb{R}^{*}$. In view of \eqref{tildWN}  and by using \eqref{usefetl}, one has:
\begin{equation*}
\Vert(\Omega_{L}(t_{1},s) - \Omega_{L}(t_{2},s)) \varphi\Vert_{2} \leq  \int_{t_{1}}^{t_{2}} \mathrm{d}\tau\, \vert \mathpzc{a}(\tau)\vert \{C \Vert(H_{L,-}+ i\alpha)\varphi\Vert_{2} +  \frac{N}{2} \vert \mathpzc{a}(\tau)\vert\Vert \varphi\Vert_{2}\}.
\end{equation*}
Thus for $T> t_{2} > t_{1}$, $\Vert (\Omega_{L}(t_{1},s) - \Omega_{L}(t_{2},s)) \varphi\Vert_{2}$ goes to zero as $t_{2} \rightarrow -\infty$. Hence, $\forall s \in \mathbb{R}$ $(\Omega_{L}(\cdot\,,s) \varphi)$ is Cauchy as $t \rightarrow -\infty$. Since $\mathrm{D}(H_{L,-})$ is dense, then $\Omega_{L}(\cdot\,,s)$ is a strong Cauchy family. From \cite[Thm. 4.23]{We}, one concludes that for any $s \in \mathbb{R}$ there exists a bounded operator $\Omega_{L}^{+}(s)$ s.t. $\Omega_{L}(t,s) \rightarrow \Omega_{L}^{+}(s)$ when $t\rightarrow -\infty$ in the strong sense. From the foregoing, one cannot conclude that $\{\Omega_{L}^{+}(s),\, s \in \mathbb{R}\}$ is unitary since the strong limit of a family of unitary operators may not be unitary (but it is isometric). Therefore, it remains to prove the unitarity property.
By a similar reasoning, one can prove that for any $s\in \mathbb{R}$ there exists a bounded operator $\Omega_{L}^{*+}(s)$ s.t. $\Omega_{L}^{*}(t,s) \rightarrow \Omega_{L}^{*+}(s)$ when $t\rightarrow -\infty$ in the strong sense. Finally, by using that $\{\Omega_{L}(t,s),\, (t,s) \in \mathbb{R}^{2}\}$ is unitary:
\begin{equation*}
\Omega_{L}^{+*}(s) \Omega_{L}^{+}(s) = \mathpzc{s}-\lim_{t \rightarrow -\infty} \Omega_{L}^{*}(t,s) \Omega_{L}(t,s) = \mathbbm{1}. \tag*{\qed}
\end{equation*}

\noindent \textbf{Proof of Lemma \ref{preservD}.} Let $s, t \in \mathbb{R}$ and $\alpha \in \mathbb{R}^{*}$. Let $\phi \in \mathrm{D}(H_{L,-})$ and $\psi \in \mathfrak{h}_{L,-}^{N}$. Define:
\begin{equation*}
\begin{split}
\ell_{s,t}(\phi,\psi) :&= \langle (H_{L,-}-i\alpha) \phi, \Omega_{L}(t,s)(H_{L,-}+i\alpha)^{-1} \psi \rangle \\
&= \langle(H_{L,-} - i\alpha) \mathrm{e}^{-i(t-s)H_{L,-}} \phi, U_{L}(t,s) (H_{L,-}+i\alpha)^{-1} \psi \rangle,
\end{split}
\end{equation*}
where we used the definition \eqref{defOmega} of $\Omega_{L}(t,s)$ in the last equality. Since:
\begin{multline*}
U_{L}(t,s) (H_{L,-}+i\alpha)^{-1} = \\ (H_{L,-}(t) + i\alpha)^{-1}(H_{L,-}(t) + i\alpha)U_{L}(t,s) (H_{L,-}(t)+i\alpha)^{-1}(H_{L,-}(t)+i\alpha) (H_{L,-}+i\alpha)^{-1},
\end{multline*}
then by the Cauchy-Schwartz inequality:
\begin{multline*}
\vert \ell_{s,t}(\phi,\psi)\vert \leq  \Vert (H_{L,-}+i\alpha)(H_{L,-}(t) + i\alpha)^{-1}\Vert \Vert (H_{L,-}(t) + i\alpha) U_{L}(t,s) (H_{L,-}(t)+i\alpha)^{-1}\Vert \\
\times \Vert (H_{L,-}(t)+i\alpha)(H_{L,-}+i\alpha)^{-1}\Vert \Vert \phi\Vert_{2} \Vert \psi\Vert_{2}.
\end{multline*}
Let $T \geq 0$ s.t. $s,t \leq T$. From Lemmas \ref{estimnorm1} and \ref{kyest} there exists $C>0$ s.t. $\forall E \in [-1,1]$:
\begin{equation*}
\vert \ell_{s,t}(\phi,\psi)\vert \leq C  \Vert \phi\Vert_{2} \Vert \psi\Vert_{2}.
\end{equation*}
The B.L.T. theorem allows us to conclude. The other estimates follow by similar arguments. \qed

\subsection{Proof of Lemmas \ref{lem30}--\ref{lem31}.}
\label{Sec32}

For simplicity's sake, we drop the dependence in $\omega$ and $\eta$ in our notation. \\

\noindent \textbf{\textit{Proof of Lemma \ref{lem30}}}. Let $\alpha \in \mathbb{R}^{*}$. We start with $\mathrm{(i)}$. We use the two following rewritings:
\begin{equation*}
P_{L,-}^{k} \rho_{L}^{\mathrm{eq}}(\beta) P_{L,-}^{l} = P_{L,-}^{k}(H_{L,-}+i \alpha)^{-1} (H_{L,-}+i\alpha) \rho_{L}^{\mathrm{eq}}(\beta) (H_{L,-}+i\alpha) (H_{L,-}+i\alpha)^{-1} P_{L,-}^{l},\\
\end{equation*}
\begin{multline}
\label{gbfvxc}
[\rho_{L}^{\mathrm{eq}}(\beta),P_{L,-}] \mathrm{e}^{-it H_{L,-}} P_{L,-} =
\rho_{L}^{\mathrm{eq}}(\beta)(H_{L,-}+i\alpha)(H_{L,-}+i\alpha)^{-1} P_{L,-} \mathrm{e}^{-it H_{L,-}} P_{L,-} +\\
- P_{L,-}(H_{L,-}+i\alpha)^{-1} (H_{L,-}+i\alpha) \rho_{L}^{\mathrm{eq}}(\beta) (H_{L,-}+i\alpha) \mathrm{e}^{-it H_{L,-}} (H_{L,-}+i\alpha)^{-1} P_{L,-}.
\end{multline}
Due to the $*$-ideal property of $\mathfrak{I}_{1}$, it suffices to use that  $(H_{L,-}+i\alpha)^{-1} P_{L,-} \mathrm{e}^{-it H_{L,-}} P_{L,-}$ and $(H_{L,-}+i\alpha)^{-1}P_{L,-}$ are bounded by Lemma \ref{relbord}, and $(H_{L,-}+i\alpha) \rho_{L}^{\mathrm{eq}}(\beta) (H_{L,-}+i\alpha)$ is trace-class by Lemma \ref{trass}. To prove $\mathrm{(ii)}$, we use the rewriting of $\rho_{L}(\beta;E,t)= \rho_{L}(\beta;t)$ given in \eqref{rewrhot}.
Recall that all operators between braces are bounded and their operator norm can be bounded uniformly in $t\in (-\infty,T]$, see  Lemma \ref{preservD}. Since $(H_{L,-}+i\alpha)^{-1}P_{L,-}$ is bounded and $(H_{L,-}+i\alpha) \rho_{L}^{\mathrm{eq}}(\beta) (H_{L,-}+i\alpha)$ is trace-class, then $P_{L,-}^{l}\rho_{L}(\beta;E,t)P_{L,-}^{k} \in \mathfrak{I}_{1}(\mathfrak{h}_{L,-}^{N})$. For the second part, we use an identity similar to \eqref{gbfvxc} but with $\rho_{L}(\beta;E,t)$ instead of $\rho_{L}^{\mathrm{eq}}(\beta)$. Then use that $\rho_{L}(\beta;E,t)(H_{L,-}+i\alpha)$ and $(H_{L,-}+i\alpha)\rho_{L}(\beta;E,t)(H_{L,-}+i\alpha)$ are trace-class (can be read off from \eqref{rewrhot}), and their trace-norm are bounded uniformly in $t\leq T$. \qed \\

\noindent \textbf{Proof of Lemma \ref{prop31}.} Due to Lemma \ref{lem30}, it suffices to prove the following identity on $\mathfrak{I}_{1}(\mathfrak{h}_{L,-}^{N})$:
\begin{equation}
\label{rhoLnew}
\rho_{L}(\beta;E,t) = \rho_{L}^{\mathrm{eq}}(\beta) - i \frac{e}{m_{e}} E \int_{-\infty}^{t} \mathrm{d}\tau\, \mathpzc{a}(\tau) \mathrm{e}^{i(\tau-t) H_{L,-}}[\rho_{L}(\beta;E,\tau),P_{L,-}]\mathrm{e}^{-i(\tau-t) H_{L,-}}.
\end{equation}
To do that, let us define the following family of trace-class operators:
\begin{equation}
\label{rhotild}
\widetilde{\rho}_{L}(\beta;E,t) := \mathrm{e}^{it H_{L,-}} \rho_{L}(\beta;E,t) \mathrm{e}^{-itH_{L,-}} = \Omega_{L}(t,0)\Omega_{L}^{*}(0) \rho_{L}^{\mathrm{eq}}(\beta) \Omega_{L}^{+}(0)\Omega_{L}^{*}(t,0),\quad t\in\mathbb{R}.
\end{equation}
In view of \eqref{tildWN} and by mimicking the arguments used in the proof of Proposition \ref{soluLoui}, we prove:
\begin{equation*}
\lim_{h \rightarrow 0} \Vert \frac{\widetilde{\rho}_{L}(\beta;E,t_{0}+h) - \widetilde{\rho}_{L}(\beta;E,t_{0})}{h} - i[\widetilde{\rho}_{L}(\beta;E,t_{0}), \widetilde{W}_{L,-}(t_{0},0)]\Vert_{1} = 0.
\end{equation*}
Since $\lim_{t \rightarrow -\infty} \Vert \widetilde{\rho}_{L}(\beta;E,t) - \rho_{L}^{\mathrm{eq}}(\beta) \Vert_{1} \leq \lim_{t \rightarrow -\infty} \Vert \rho_{L}(\beta;E,t) - \rho_{L}^{\mathrm{eq}}(\beta) \Vert_{1} = 0$ by \eqref{inicondi}, one has:
\begin{equation*}
\widetilde{\rho}_{L}(\beta;E,t) = \rho_{L}^{\mathrm{eq}}(\beta) + i \int_{-\infty}^{t} \mathrm{d}\tau\, [\widetilde{\rho}_{L}(\beta;E,\tau), \widetilde{W}_{L,-}(\tau,0)].
\end{equation*}
\eqref{rhoLnew} follows from $[\widetilde{\rho}_{L}(\beta;E,\tau), \widetilde{W}_{L,-}(\tau,0)]=  \mathrm{e}^{i\tau H_{L,-}} [\rho_{L}(\beta;E,\tau), W_{L,-}(\tau)] \mathrm{e}^{-i\tau H_{L,-}}$ along with \eqref{defW}. \qed \\

Let us turn to the proof of Lemma \ref{prop32}. To do that, we need the following estimate:

\begin{lema}
\label{lemaad3}
$\forall L \in \mathbb{N}^{*}$, $\forall \omega>0$, $\forall \eta>0$ and $\forall \alpha \in \mathbb{R}^{*}$ there exists $C>0$ s.t. $\forall E \in [-1,1]$:
\begin{equation}
\label{alda1}
\sup_{t \in (-\infty,0]} \Vert [\mathbbm{1} - \Omega_{L}(t,0) \Omega_{L}^{+*}(0)](H_{L,-}+i\alpha)^{-1}\Vert \leq C \vert E\vert (1+\vert E\vert).
\end{equation}
\end{lema}

\noindent \textbf{\textit{Proof of Lemma \ref{lemaad3}}.} For any $t \leq 0$, one has in the bounded operators sense:
\begin{equation*}
[\mathbbm{1} - \Omega_{L}(t,0) \Omega_{L}^{+*}(0)](H_{L,-}+i\alpha)^{-1} =
[\mathbbm{1} - \Omega_{L}(t,0)]\Omega_{L}^{+*}(0) (H_{L,-}+i\alpha)^{-1}  + [\mathbbm{1} - \Omega_{L}^{+*}(0)] (H_{L,-}+i\alpha)^{-1}.
\end{equation*}
Let $\psi \in \mathfrak{h}_{L,-}^{N}$. From \eqref{relOM}, the first term on the r.h.s. can be rewritten as:
\begin{multline*}
[\mathbbm{1} - \Omega_{L}(t,0)]\Omega_{L}^{+*}(0) (H_{L,-}+i\alpha)^{-1}\psi =
- i  (\int_{t}^{0} \mathrm{d}\tau\, \mathrm{e}^{i\tau H_{L,-}} W_{L,-}(\tau) (H_{L,-}+i\alpha)^{-1} \mathrm{e}^{-i\tau H_{L,-}} \\
\times \{(H_{L,-}+i\alpha) \Omega_{L}(\tau,0) (H_{L,-}+i\alpha)^{-1}\} \{(H_{L,-}+i\alpha) \Omega_{L}^{+*}(0)(H_{L,-} + i\alpha)^{-1}\})\psi.
\end{multline*}
To treat the second term on the r.h.s., we use again \eqref{relOM}:
\begin{equation*}
[\mathbbm{1} - \Omega_{L}^{*}(t,0)] (H_{L,-}+i\alpha)^{-1}\psi = i(\int_{t}^{0} \mathrm{d}\tau\, \Omega_{L}^{*}(\tau,0) \mathrm{e}^{i \tau H_{L,-}} W_{L,-}(\tau) (H_{L,-}+i\alpha)^{-1} \mathrm{e}^{-i\tau H_{L,-}})\psi.
\end{equation*}
From \eqref{normPjres} along with \eqref{est11}-\eqref{est12}, there exists $C>0$ s.t. $\forall E \in [-1,1]$ and $\forall t \in (-\infty,0]$:
\begin{multline*}
\max\{\Vert [\mathbbm{1} - \Omega_{L}(t,0)]\Omega_{L}^{+*}(0) (H_{L,-}+i\alpha)^{-1} \Vert, \Vert [\mathbbm{1} - \Omega_{L}^{*}(t,0)] (H_{L,-}+i\alpha)^{-1}\Vert \} \\
\leq  C \vert E\vert \int_{-\infty}^{0} \mathrm{d}\tau\, \vert \mathpzc{a}(\tau) \vert ( 1 + N \vert E\vert \vert \mathpzc{a}(\tau)\vert).
\end{multline*}
As a result of Lemma \ref{stron}, such an estimate also holds true for $\Vert [\mathbbm{1} - \Omega_{L}^{*+}(0)] (H_{L,-}+i\alpha)^{-1} \Vert$. \qed \\

\noindent \textbf{Proof of Lemma \ref{prop32}.} We start with the case of $k=1$. From \eqref{R2}:
\begin{multline*}
\mathrm{Tr}_{\mathfrak{h}_{L,-}^{N}}\{\mathscr{R}_{L}(\beta;E,0)P_{L,-}\} = \\ i  \int_{-\infty}^{0} \mathrm{d}\tau\, \mathpzc{a}(\tau) \mathrm{Tr}_{\mathfrak{h}_{L,-}^{N}}\{\mathrm{e}^{i\tau H_{L,-}}[P_{L,-}, (\rho_{L}(\beta;E,\tau) - \rho_{L}^{\mathrm{eq}}(\beta))]\mathrm{e}^{- i \tau H_{L,-}} P_{L,-}\}.
\end{multline*}
By introducing the operator $Q_{L,-} := \mathrm{e}^{-i\tau H_{L,-}} P_{L,-} \mathrm{e}^{i\tau H_{L,-}} P_{L,-}$, one has by cyclicity of the trace:
\begin{multline*}
\mathrm{Tr}_{\mathfrak{h}_{L,-}^{N}}\{\mathscr{R}_{L}(\beta;E,0)P_{L,-}\} = \\ i \int_{-\infty}^{0} \mathrm{d}\tau\, \mathpzc{a}(\tau) \mathrm{Tr}_{\mathfrak{h}_{L,-}^{N}}\{Q_{L,-} (\rho_{L}(\beta;E,\tau) - \rho_{L}^{\mathrm{eq}}(\beta)) - (\rho_{L}(\beta;E,\tau) - \rho_{L}^{\mathrm{eq}}(\beta))Q_{L,-}^{*}\}.
\end{multline*}
Note that the above quantity is well-defined since $Q_{L,-}(H_{L,-}+i)^{-1}$ is bounded due to \eqref{newsno} and $(H_{L,-}+i)(\rho_{L}(\beta;E,\tau) - \rho_{L}^{\mathrm{eq}}(\beta))(H_{L,-}+i)$ is trace-class, see \eqref{rewrhot}  with Lemma \ref{trass}. From \eqref{rhot}:
\begin{multline*}
\mathrm{Tr}_{\mathfrak{h}_{L,-}^{N}}\{Q_{L,-} (\rho_{L}(\beta;E,\tau) - \rho_{L}^{\mathrm{eq}}(\beta))\}
=  \mathrm{Tr}_{\mathfrak{h}_{L,-}^{N}}\{Q_{L,-} \mathrm{e}^{-i\tau H_{L,-}}[\Omega_{L}(\tau,0) \Omega_{L}^{+*}(0) - \mathbbm{1}] \rho_{L}^{\mathrm{eq}}(\beta) \mathrm{e}^{i\tau H_{L,-}}\} \\
+ \mathrm{Tr}_{\mathfrak{h}_{L,-}^{N}}\{Q_{L,-} \mathrm{e}^{-i\tau H_{L,-}} \Omega_{L}(\tau,0) \Omega_{L}^{+*}(0) \rho_{L}^{\mathrm{eq}}(\beta)[\Omega_{L}^{+}(0) \Omega_{L}^{*}(\tau,0) - \mathbbm{1}] \mathrm{e}^{i\tau H_{L,-}}\}.
\end{multline*}
Let $\alpha \in \mathbb{R}^{*}$. By cyclicity of the trace for the first term on the r.h.s., one gets the upper bound:
\begin{multline*}
\Vert \mathrm{e}^{-i\tau H_{L,-}}[\Omega_{L}(\tau,0) \Omega_{L}^{+*}(0) - \mathbbm{1}] \rho_{L}^{\mathrm{eq}}(\beta) \mathrm{e}^{i\tau H_{L,-}} Q_{L,-} \Vert_{1} \\
\leq \Vert [\Omega_{L}(\tau,0) \Omega_{L}^{+*}(0) - \mathbbm{1}] (H_{L,-}+i\alpha)^{-1}\Vert \Vert  (H_{L,-}+i\alpha) \rho_{L}^{\mathrm{eq}}(\beta)  (H_{L,-}+i\alpha)\Vert_{1} \Vert (H_{L,-}+i\alpha)^{-1} Q_{L,-} \Vert,
\end{multline*}
as for the second term:
\begin{multline*}
\Vert Q_{L,-} \mathrm{e}^{-i \tau H_{L,-}} \Omega_{L}(\tau,0)\Omega_{L}^{+*}(0)\rho_{L}^{\mathrm{eq}}(\beta) \{\Omega_{L}^{+}(0)\Omega_{L}^{*}(\tau,0) - \mathbbm{1}\} \mathrm{e}^{i\tau H_{L,-}}\Vert_{1} \\
\leq \Vert Q_{L,-} (H_{L,-}+i\alpha)^{-1}\Vert \Vert (H_{L,-}+i\alpha) \Omega_{L}(\tau,0) (H_{L,-}+i\alpha)^{-1} \Vert \Vert (H_{L,-}+i\alpha)\Omega_{L}^{+*}(0)(H_{L,-}+i\alpha)^{-1}\Vert \\
\times \Vert (H_{L,-}+i\alpha)\rho_{L}^{\mathrm{eq}}(\beta) (H_{L,-}+i\alpha)\Vert_{1} \Vert (H_{L,-}+i\alpha)^{-1}\{\Omega_{L}^{+}(0)\Omega_{L}^{*}(\tau,0) - \mathbbm{1}\}\Vert.
\end{multline*}
By using \eqref{newsno} and Lemmas \ref{trass}, \ref{preservD} and \ref{lemaad3} then there exists $C>0$ s.t. $\forall E \in [-1,1]$:
\begin{equation*}
\sup_{\tau \in (-\infty,0]}\Vert Q_{L,-}  (\rho_{L}(\beta;E,\tau) - \rho_{L}^{\mathrm{eq}}(\beta))\Vert_{1} \leq C \vert E \vert (1 + \vert E \vert).
\end{equation*}
Note that the same upper bound holds true for $\Vert(\rho_{L}(\beta;E,\tau) - \rho_{L}^{\mathrm{eq}}(\beta)) Q_{L,-}^{*}\Vert_{1}$. \eqref{labon} follows from:
\begin{equation*}
\forall E \in [-1,1],\quad \vert \mathrm{Tr}_{\mathfrak{h}_{L,-}^{N}}\{\mathscr{R}_{L}(\beta;E,t=0)P_{L,-}\} \vert \leq C \vert E\vert(1+ \vert E\vert) \int_{-\infty}^{0} \mathrm{d}\tau \vert \mathpzc{a}(\tau)\vert,
\end{equation*}
for another constant $C>0$. Let us turn to the second part of the Lemma. Firstly from \eqref{R2}, $\mathrm{Tr}_{\mathfrak{h}_{L,-}^{N}}\{\mathscr{R}_{L}(\beta;0,0)P_{L,-}\} = 0$
since $\rho_{L}(\beta;E=0,t=0) = \rho_{L}^{\mathrm{eq}}(\beta)$. Secondly, for $\vert h\vert <1$ the quantity:
\begin{equation*}
\vert i \int_{-\infty}^{0} \mathrm{d}\tau\, \mathpzc{a}(\tau)  \mathrm{Tr}_{\mathfrak{h}_{L,-}^{N}} \{\mathrm{e}^{i\tau H_{L,-}}[P_{L,-}, (\rho_{L}(\beta;h,\tau) - \rho_{L}^{\mathrm{eq}}(\beta))] \mathrm{e}^{-i\tau H_{L,-}}P_{L,-}\}\vert,
\end{equation*}
is bounded above by $const \times \vert h \vert$ by \eqref{labon} and then admits zero as limit when $h\rightarrow 0$. Therefore,
$E \mapsto E \mathrm{Tr}_{\mathfrak{h}_{L,-}^{N}}\{\mathscr{R}_{L}(\beta;E,0)P_{L,-}\}$ is differentiable at $E=0$ with derivative equal to zero.\\
The case of $k=0$ can be treated by similar arguments. \qed \\

\noindent \textbf{Proof of Lemma \ref{lem31}.} In view of formula \eqref{T1}, we have the upper-bound:
\begin{equation*}
\vert \mathpzc{T}_{L}^{(1)}(\beta)\vert \leq 2 \sum_{k=1}^{\infty} \mathrm{e}^{-\beta(\mu_{k} - \mu_{0})}\Vert P_{L,-} \psi_{k} \Vert_{2}^{2} \int_{-\infty}^{0} \mathrm{d}\tau\, \vert \mathpzc{a}(\tau)\vert,
\end{equation*}
where we used the Cauchy-Schwartz inequality and the fact that $\tilde{Z}_{\beta,L} \geq 1$ by virtue of \eqref{partmod}. By using \eqref{fdstr} and \eqref{defa}, there exist $c_{1}=c_{1}(\eta,\omega)>0$ and (another) $c_{2}=c_{2}(N)>0$ s.t.
\begin{equation*}
\vert \mathpzc{T}_{L}^{(1)}(\beta)\vert \leq c_{1} N \sum_{k=1}^{\infty} (\mu_{k} + c_{2}) \mathrm{e}^{-\beta(\mu_{k} - \mu_{0})}.
\end{equation*}
It remains to estimate the above series. From \eqref{asymptotf} we infer that there exists $k_{0} \in \mathbb{N}^{*}$ s.t. $\forall k \geq k_{0}$:
\begin{equation*}
\mathpzc{C}_{0}  k^{\frac{2}{N}} \leq 2 \mu_{k} \leq 3 \mathpzc{C}_{0} k^{\frac{2}{N}},
\end{equation*}
with $\mathpzc{C}_{0}>0$ the constant in \eqref{asymptotf}. Moreover, there exists $k_{1} \in \mathbb{N}^{*}$ s.t. $\forall k \geq k_{1}$, $\mathpzc{C}_{0} k^{\frac{2}{N}} \leq 4(\mu_{k} - \mu_{0})$. Denoting by $\varkappa := \max\{k_{0},k_{1}\} \in \mathbb{N}^{*}$, we then obtain:
\begin{equation*}
\sum_{k=1}^{\infty} (\mu_{k} + c_{2}) \mathrm{e}^{-\beta(\mu_{k} - \mu_{0})} \leq \mathrm{e}^{-\beta(\mu_{1} - \mu_{0})} \sum_{k=1}^{\varkappa+1} (\mu_{k}+c_{2}) + (\frac{3 \mathpzc{C}_{0}}{2} + c_{2}) \int_{\varkappa}^{\infty} \mathrm{d}s\,  (s^{\frac{2}{N}} + 1) \mathrm{e}^{-\frac{\mathpzc{C}_{0}}{4} \beta s^{\frac{2}{N}}}.
\end{equation*}
Note that the integrals on the r.h.s. can be expressed in terms of the incomplete Gamma function, see \cite[Eq. (6.5.3)]{AS}. Using its asymptotic behavior in \cite[Eq. (6.5.32)]{AS}, the second term on the r.h.s can be bounded for $\beta$ sufficiently large by $c_{3} \mathrm{e}^{-c_{4} \beta}/\beta$ for some $c_{l}=c_{l}(N)>0$, $l \in \{3,4\}$. \qed

\section{Appendix: Complementary results and missing proofs.}

\subsection{A few technical results.}
\label{Serrest}

Here, we collect some properties on the family of operators introduced in \eqref{defHN}. We also give a series of useful estimates on operator norms. The proofs of the Lemmas below lie in Sec. \ref{restprv}.\\

Let $\{\mu_{k}\}_{k\geq 0}$, $\mu_{k}=\mu_{k}(L,N)$ with $N=N_{L}$ obeying \eqref{relsemic} be the set of eigenvalues of $H_{L,-}$ counting multiplicities and in increasing order. They satisfy the following asymptotic:

\begin{lema}
\label{asymptot}
There exists a constant $\mathpzc{C}_{0}=\mathpzc{C}_{0}(L,N)>0$ s.t.
\begin{equation}
\label{asymptotf}
\mu_{k} \sim \mathpzc{C}_{0} k^{\frac{2}{N}} \quad \textrm{when $k \rightarrow \infty$}.
\end{equation}
\end{lema}

From Lemma \ref{asymptot}, we have the following:

\begin{lema}
\label{trass}
$\forall L \in \mathbb{N}^{*}$, $\forall \beta>0$ and $\forall l \in \mathbb{N}^{*}$, $H_{L,-}^{l} \mathrm{e}^{-\beta H_{L,-}}$ is a trace-class operator on $\mathfrak{h}_{L,-}^{N}$.
\end{lema}

We next turn to a series of estimates on operator norms:

\begin{lema}
\label{relbord}
$\mathrm{(i)}$. $\forall L \in \mathbb{N}^{*}$ and $\forall \alpha \in \mathbb{R}^{*}$ there exists $C>0$ s.t.
\begin{equation}
\label{normPjres}
\Vert P_{L,-} (H_{L,-} + i\alpha)^{-1} \Vert \leq C.
\end{equation}
$\mathrm{(ii)}$. $\forall L \in \mathbb{N}^{*}$ and $\forall \alpha \in \mathbb{R}^{*}$  there exists $C>0$ s.t. $\forall t \in \mathbb{R}$:
\begin{equation}
\label{newsno}
\Vert P_{L,-} \mathrm{e}^{-it H_{L,-}} P_{L,-} (H_{L,-}+i\alpha)^{-1}\Vert \leq C.
\end{equation}
\end{lema}

\begin{lema}
\label{estimnorm1}
$\mathrm{(i)}$. $\forall L \in \mathbb{N}^{*}$, $\forall \omega>0$, $\forall \eta>0$, $\forall T\geq 0$, $\forall \alpha \in \mathbb{R}^{*}$ there exists $C>0$ s.t. $\forall E \in [-1,1]$:
\begin{equation}
\label{norm1}
\sup_{t \in (-\infty,T]} \Vert W_{L,-}(t) (H_{L,-} + i \alpha)^{-1}\Vert \leq C.
\end{equation}
$\mathrm{(ii)}$. $\forall L \in \mathbb{N}^{*}$, $\forall \omega>0$, $\forall \eta>0$, $\forall T\geq 0$ and $\forall \alpha \in \mathbb{R}^{*}$ there exists $C>0$ s.t. $\forall E \in [-1,1]$:
\begin{equation}
\label{norm2}
\sup_{t \in (-\infty,T]} \Vert (H_{L,-} + i \alpha) (H_{L,-}(t) + i \alpha)^{-1}\Vert \leq C.
\end{equation}
\end{lema}

\subsection{The missing proofs.}

\subsubsection{Construction of the family of operators \eqref{HLr}.}
\label{constutt}

Define the non-negative symmetric sesquilinear form $h_{L,r}^{(C)}: \mathcal{C}^{\infty}(\mathcal{T}_{La,r}^{N})\times \mathcal{C}^{\infty}(\mathcal{T}_{La,r}^{N})\rightarrow \mathbb{C}$ by:
\begin{equation}
\label{form1}
h_{L,r}^{(C)}(\phi,\psi) := \frac{1}{2} \sum_{j=1}^{N} (\langle -i \partial_{x_{j}} \phi , - i\partial_{x_{j}} \psi\rangle + \langle -i \partial_{y_{j}} \phi, - i\partial_{y_{j}} \psi\rangle)  + \frac{\lambda}{2}\sum_{j\neq l = 1}^{N} \langle \sqrt{V_{L,r}} \phi, \sqrt{V_{L,r}} \psi \rangle.
\end{equation}
The first term on the r.h.s. is the 'kinetic' sesquilinear form whose closure has domain the Sobolev space $\mathcal{W}^{1,2}(\mathcal{T}_{La,r}^{N})$. We denote by $\mathcal{H}_{L,r}^{(0)}:=  \frac{1}{2}\sum_{j=1}^{N} (- \Delta_{x_{j}} - \Delta_{y_{j}})$ its associated self-adjoint operator. Note that the second term is well-defined due to Lemma \ref{propVcL} $\mathrm{(i)}$-$\mathrm{(ii)}$. It is the sesquilinear form associated to the periodized Coulomb potential energy whose maximal domain is:
\begin{equation*}
Q:= \{\varphi \in L^{2}(\mathcal{T}_{La,r}^{N}): \sum_{ j\neq l = 1}^{N} \int_{\mathcal{T}_{La,r}^{N}} \mathrm{d}\bold{x}\mathrm{d}\bold{y}\, \vert \sqrt{V_{L,r}(x_{j}-x_{l},y_{j}-y_{l})} \varphi(\bold{x},\bold{y})\vert^{2} < \infty\}.
\end{equation*}
Extended to $\mathcal{Q}(h_{L,r}^{(C)}) := Q \cap \mathcal{W}^{1,2}(\mathcal{T}_{La,r}^{N})$, the form \eqref{form1} is densely-defined in $L^{2}(\mathcal{T}_{La,r}^{N})$, non-negative, symmetric and closed. By \cite[Thm. VIII.15]{RS1}, it generates a unique positive self-adjoint operator with form core $\mathcal{C}^{\infty}(\mathcal{T}_{La,r}^{N})$. We denote it by $\mathcal{H}_{L,r}^{(C)}$ and represent it for convenience as:
\begin{equation*}
\mathcal{H}_{L,r}^{(C)} := \mathcal{H}_{L,r}^{(0)} + \frac{\lambda}{2} \sum_{j \neq l = 1}^{N} V_{L,r}(x_{j}-x_{l},y_{j}-y_{l}).
\end{equation*}
Due to our assumption on $V_{\mathrm{per}}$, the sesquilinear form associated to the electric potential energy  is infinitesimally form-bounded relative to $\mathcal{H}_{L,r}^{(0)}$, and then to $\mathcal{H}_{L,r}^{(C)}$ since $\mathcal{H}_{L,r}^{(0)}\leq \mathcal{H}_{L,r}^{(C)}$. By the KLMN Theorem \cite[Thm X.17]{RS2}, the sesquilinear form $h_{L,r}: \mathcal{Q}(h_{L,r}^{(C)})\times \mathcal{Q}(h_{L,r}^{(C)})\rightarrow \mathbb{C}$ defined as:
\begin{equation*}
h_{L,r}(\phi,\psi) := h_{L,r}^{(C)}(\phi,\psi) + \sum_{j=1}^{N} \langle \phi , V_{\mathrm{per}} \psi\rangle,
\end{equation*}
is closed, bounded from below and with form core $\mathcal{C}^{\infty}(\mathcal{T}_{La,r}^{N})$. By \cite[Thm. VIII.15]{RS1}, it generates a unique bounded from below self-adjoint operator on $L^{2}(\mathcal{T}_{La,r}^{N})$. We represent it as \eqref{HLr}. Note that the above construction of $\mathcal{H}_{L,r}$ corresponds to impose periodic boundary conditions on $[-\frac{La}{2},\frac{La}{2})^{N}$.

\subsubsection{Proof of Lemmas \ref{propVcL}--\ref{propvL}.}
\label{PropPoten}

For simplicity's sake, we set $a=1$ in the definition \eqref{defVcrL} and $\varepsilon=1=e$ in \eqref{Vcrdef}.\\

\noindent \textbf{Proof of Lemma \ref{propVcL}}. Let $L \in \mathbb{N}^{*}$ and $0<2\sqrt{2}r<1$.  For any $x\in \mathbb{R}$ and $y\in [-\pi r,0[\cup]0,\pi r[$:
\begin{gather}
\label{rewwwd}
V_{L,r}(x,y) =I_{L,1}(x,y) + I_{L,2}(x,y),\quad \textrm{with:}  \\
I_{L,1}(x,y) := \frac{1}{L} \sum_{m\in \mathbb{Z}} \mathrm{e}^{i \frac{2\pi}{L} m x} \int_{-\frac{L}{2}}^{\frac{L}{2}} \mathrm{d}x'\, \mathrm{e}^{-i \frac{2\pi}{L}m x'} V_{r}(x',y),\nonumber\\
\label{I2}
I_{L,2}(x,y) := \frac{1}{L}\sum_{m\in \mathbb{Z}^{*}} \mathrm{e}^{i \frac{2\pi}{L} m x} \int_{\frac{L}{2}}^{\infty} \mathrm{d}x'\, 2 \cos(\frac{2\pi}{L}m x') V_{r}(x',y).
\end{gather}
Let us first prove that $I_{L,2}$ is uniformly bounded. On the one hand, \eqref{I2} can be rewritten as:
\begin{equation*}
I_{L,2}(x,y) = \frac{4}{L} \sum_{m=1}^{\infty} \cos(\frac{2\pi}{L} m x) J_{L}(m,y),\quad J_{L}(m,y):= \int_{\frac{L}{2}}^{\infty} \mathrm{d}x'\, \cos(\frac{2\pi}{L} m x') V_{r}(x',y).
\end{equation*}
On the other hand, the first four derivatives of $V_{r}(\cdot\,,y)$ read as:
\begin{gather}
\label{dervVcr}
(\partial_{x} V_{r})(x,y) = \frac{-x}{(x^{2} + 4r^{2}\sin^{2}(\frac{y}{2r}))^{\frac{3}{2}}}, \quad
(\partial_{x}^{2} V_{r})(x,y) = \frac{2x^{2} - 4r^{2}\sin^{2}(\frac{y}{2r})}{(x^{2} + 4r^{2}\sin^{2}(\frac{y}{2r}))^{\frac{5}{2}}},\\
(\partial_{x}^{3} V_{r})(x,y) = \frac{6(6 r^{2}\sin^{2}(\frac{y}{2r})- x^{2}) x}{(x^{2} + 4r^{2}\sin^{2}(\frac{y}{2r}))^{\frac{7}{2}}}, \quad (\partial_{x}^{4} V_{r})(x,y) = \frac{24(6 r^{4}\sin^{4}(\frac{y}{2r}) - 12 r^{2}\sin^{2}(\frac{y}{2r}) x^{2} + x^{4})}{(x^{2} + 4r^{2}\sin^{2}(\frac{y}{2r}))^{\frac{9}{2}}}.\nonumber
\end{gather}
Note that $(\partial_{x}^{l} V_{r})(\cdot\,,y) \in L^{1}(\mathbb{R})$, $l\in \{1,2,3,4\}$ since $y\neq 0$ and $\lim_{\vert x \vert \rightarrow \infty} \vert (\partial_{x}^{l} V_{r})(x,y)\vert=0$.\\
From the foregoing, one has by successive integrations by parts:
\begin{equation}
\label{JLm}
J_{L}(m,y) = - \cos(\pi m) (\frac{L}{2\pi m})^{2} (\partial_{x} V_{r})(\frac{L}{2},y) - (\frac{L}{2\pi m})^{2} \int_{\frac{L}{2}}^{\infty} \mathrm{d}x'\, \cos(\frac{2\pi}{L} m x')(\partial_{x}^{2} V_{r})(x',y).
\end{equation}
Note that $(\partial_{x}^{2} V_{r})(\cdot\,,y) \geq 0$ on $[\frac{L}{2},\infty)$ since $L \geq 1 \geq 2\sqrt{2}r$ by assumption. Therefore:
\begin{equation*}
\vert J_{L}(m,y)\vert \leq 2 (\frac{L}{2\pi m})^{2} [-(\partial_{x} V_{r})(\frac{L}{2},y)].
\end{equation*}
From the definition of $I_{L,2}$ in \eqref{I2} along with the expression of $(\partial_{x} V_{r})(\cdot\,,y)$, one arrives at:
\begin{equation}
\label{uppI2}
\forall (x,y) \in \mathbb{R}\times ([-\pi r,0[\cup]0,\pi r[),\quad \vert I_{L,2}(x,y) \vert \leq  \frac{L}{3} [-(\partial_{x} V_{r})(\frac{L}{2},y)].
\end{equation}
Now we can start the actual proof of Lemma \ref{propVcL}. Since $I_{L,1}(\cdot\,,y)$ corresponds to the complete Fourier periodic expansion of $V_{r}(\cdot\,,y)$ restricted to the interval $[-\frac{L}{2},\frac{L}{2})$, then one has:
\begin{equation}
\label{fdesdA}
\forall x \in [-\frac{L}{2},\frac{L}{2}),\quad V_{L,r}(x,y) = V_{r}(x,y) + I_{L,2}(x,y),\quad y \neq 0.
\end{equation}
By using that $I_{L,2}$ is uniformly bounded by $const/L$ (see \eqref{uppI2} with \eqref{dervVcr}) and $V_{r} \in L^{1}_{\mathrm{loc}}(\mathcal{C}_{\infty,r})$, then $\mathrm{(i)}$ follows. $\mathrm{(iii)}$ is straightforward since $\vert I_{L,2}(x,y)\vert = \mathcal{O}(L^{-1})$ when $L \rightarrow \infty$. Let us turn to $\mathrm{(ii)}$. Since $V_{r}(\cdot\,,y) \upharpoonright [-\frac{L}{2},\frac{L}{2})$ reaches its minimum at $x=-\frac{L}{2}$, then from \eqref{fdesdA} along with \eqref{uppI2}:
\begin{equation*}
V_{L,r}(x,y) \geq V_{r}(-\frac{L}{2},y) - \vert I_{L,2}(x,y)\vert \geq (1 -\frac{2}{3})\frac{1}{\sqrt{(\frac{L}{2})^{2} + 4r^{2}\sin^{2}(\frac{y}{2r})}} >0.
\end{equation*}
Moreover, from the above expressions of $(\partial_{x}^{l} V_{r})(\cdot\,,y)$ one has by successive integrations by parts:
\begin{equation*}
\int_{\mathbb{R}} \mathrm{d}x'\, \mathrm{e}^{-i\frac{2\pi}{L} mx'} V_{r}(x',y) = \frac{1}{i^{l}} \frac{L^{l}}{(2\pi m )^{l}}\int_{\mathbb{R}} \mathrm{d}x'\, \mathrm{e}^{-i\frac{2\pi}{L} mx'} (\partial_{x}^{l} V_{r})(x',y).
\end{equation*}
In view of \eqref{defVcrL} and by using that $(\partial_{x}^{l} V_{r})(\cdot\,,y) \in L^{1}(\mathbb{R})$, $l\in \{1,2,3,4\}$, then by standard arguments one gets that $x\mapsto V_{L,r}(x,y)$ is twice differentiable on $\mathbb{R}$ and its first two derivatives read as:
\begin{equation*}
(\partial_{x}^{l} V_{L,r})(x,y) = - \frac{1}{L} \sum_{m\in \mathbb{Z}^{*}} \frac{L^{2}}{(2\pi m)^{2}} \mathrm{e}^{i\frac{2\pi}{L} mx} \int_{\mathbb{R}} \mathrm{d}x'\, \mathrm{e}^{-i\frac{2\pi}{L} mx'} (\partial_{x}^{l+2} V_{r})(x',y),\quad l \in \{1,2\}.
\end{equation*}
By recursive arguments, we prove that $\mathbb{R} \owns x \mapsto V_{L,r}(x,y)$ is a $\mathcal{C}^{\infty}$-function. \qed \\

\noindent \textbf{Proof of Lemma \ref{propvL}}. We start by $\mathrm{(i)}$. From \eqref{fdesdA} together with \eqref{vrdef} and \eqref{vLdef}:
\begin{equation}
\label{fgdhty}
\forall x \in [-\frac{L}{2},0[\cup]0,\frac{L}{2}[,\quad v_{L,r}(x) = v_{r}(x) + \frac{1}{2\pi r} \int_{-\pi r}^{\pi r} \mathrm{d}y\, I_{L,2}(x,y).
\end{equation}
On the one hand, one has by virtue of \eqref{uppI2} the following upper-bound:
\begin{equation*}
\frac{1}{2\pi r} \vert \int_{-\pi r}^{\pi r} \mathrm{d}y\, I_{L,2}(x,y)\vert \leq \frac{1}{\pi r} \frac{L}{3} \int_{0}^{\pi r} \mathrm{d}y\, [-(\partial_{x} V_{r})(\frac{L}{2},y)] \leq \frac{L}{3} [-(\partial_{x} V_{r})(\frac{L}{2},0)],
\end{equation*}
which is $r$-independent. On the other hand, $v_{r} \in L^{2}(\mathbb{R})$ by virtue of \eqref{cloO1}-\eqref{cloO2}. By using that $\forall x \in \mathbb{R}$ $v_{r}(x) = r^{-1} v_{1}(x r^{-1})$,  one then obtains by the Minkowski inequality:
\begin{equation*}
\Vert v_{L,r} \Vert_{L^{2}(\mathbb{T}_{L})} \leq \frac{1}{r} \Vert v_{1}(\frac{\cdot}{r})\Vert_{L^{2}(\mathbb{T}_{L})} + \frac{L^{\frac{3}{2}}}{3} [-(\partial_{x} V_{r})(\frac{L}{2},0)] \leq \frac{1}{\sqrt{r}} (\int_{\mathbb{R}} \mathrm{d}x\, \vert v_{1}(x)\vert^{2})^{\frac{1}{2}} + \frac{4}{3}.
\end{equation*}
Let us turn to $\mathrm{(ii)}$. From \eqref{fgdhty} and by using \eqref{uppI2} along with \eqref{dervVcr}, one has $\forall x \in [-\frac{L}{2},0[\cup]0,\frac{L}{2}[$:
\begin{equation*}
v_{L,r}(x) \geq  v_{r}(x) - \frac{2}{3} \frac{1}{2\pi r} \int_{-\pi r}^{\pi r} \mathrm{d}y\, \frac{1}{\sqrt{(\frac{L}{2})^{2} + 4r^{2} \sin^{2}(\frac{y}{2r})}} = v_{r}(x) - \frac{2}{3} v_{r}(\frac{L}{2}).
\end{equation*}
It remains to use that $v_{r}$ is symmetric and decreasing on $(0,\infty)$ leading to $v_{L,r}(x) \geq \frac{1}{3} v_{r}(\frac{L}{2})>0$. We turn to $\mathrm{(iii)}$. The starting-point is \eqref{rewwwd}. Firstly, we have:
\begin{equation}
\label{FourT}
\frac{1}{2\pi r} \int_{-\pi r}^{\pi r} \mathrm{d}y\, I_{L,1}(x,y) = \frac{1}{L} \sum_{m\in \mathbb{Z}} \mathrm{e}^{i\frac{2\pi}{L} m x} \int_{-\frac{L}{2}}^{\frac{L}{2}} \mathrm{d}x'\, \mathrm{e}^{i\frac{2\pi}{L} m x'} v_{r}(x'),
\end{equation}
and since $v_{r} \in L^{2}([-\frac{L}{2},\frac{L}{2}))$ from the foregoing, then the Fourier series on the r.h.s.  converges for almost every $x$ by Carleson's theorem. Besides, since $\mathbb{R}^{*} \owns x\mapsto V_{r}(x,y)$ is smooth by Lemma \ref{propVcL} $\mathrm{(ii)}$ and $(\partial_{x}^{l} V_{r})(x,\cdot\,) \in L^{1}([-\pi r,\pi r))$, $l \in\{1,2\}$ then one has from \eqref{vrdef}:
\begin{equation*}
v_{r}^{(l)}(x) := \frac{\mathrm{d}^{l} v_{r}}{\mathrm{d}x^{l}}(x) = \frac{1}{2\pi r} \int_{-\pi r}^{\pi r} \mathrm{d}y\, (\partial_{x}^{l} V_{r})(x,y),\quad x \in \mathbb{R}^{*}.
\end{equation*}
Since $v_{r}^{(l)} \in L^{1}([\vartheta,\infty))$, $l \in \{1,2\}$ with $\vartheta>0$ then by standard arguments one obtains from \eqref{JLm}:
\begin{equation*}
\frac{1}{2\pi r} \int_{-\pi r}^{\pi r} \mathrm{d}y\, J_{L}(m,y) = - \cos(\pi m)(\frac{L}{2\pi m})^{2}v_{r}^{(1)}(\frac{L}{2}) - (\frac{L}{2\pi m})^{2} \int_{\frac{L}{2}}^{\infty} \mathrm{d}x'\, \cos(\frac{2\pi}{L} m x') v_{r}^{(2)}(x').
\end{equation*}
By successive integrations by parts one then obtains:
\begin{equation}
\label{FourT2}
\begin{split}
\frac{1}{2\pi r} \int_{-\pi r}^{\pi r} \mathrm{d}y\, I_{L,2}(x,y) &= \frac{4}{L} \sum_{m=1}^{\infty} \cos(\frac{2 \pi}{L} mx)\int_{\frac{L}{2}}^{\infty}\mathrm{d}x'\, \cos(\frac{2\pi}{L} mx') v_{r}(x') \\
&= \frac{1}{L} \sum_{m \in \mathbb{Z}^{*}} \mathrm{e}^{i\frac{2\pi}{L} m x} \int_{\vert x'\vert \geq \frac{L}{2}} \mathrm{d}x'\, \mathrm{e}^{-i\frac{2\pi}{L} m x'} v_{r}(x').
\end{split}
\end{equation}
Gathering \eqref{FourT} and \eqref{FourT2} together, the proof of $\mathrm{(iii)}$ is over. \qed

\subsubsection{Proof of Lemmas \ref{asymptot}--\ref{estimnorm1}.}
\label{restprv}

\noindent \textbf{Proof of Lemma \ref{asymptot}}. Since the set of eigenvalues $\{\mu_{k}\}_{k\geq0}$ of $H_{L,-}$ form a strict subset of the set of eigenvalues $\{\nu_{k}\}_{\geq 0}$ of $H_{L}$, it is enough to prove the asymptotic behavior for the $\nu_{k}$'s. Denote by $\{\lambda_{k}^{(\mathrm{per})}\}_{k\geq 0}$, $\{\lambda_{k}^{(\mathrm{Dir})}\}_{k\geq 0}$ and $\{\lambda_{k}^{(\mathrm{Neu})}\}_{k\geq 0}$ the set of eigenvalues of $-\frac{1}{2}\Delta_{L}$ with periodic, Dirichlet and Neumann boundary conditions on $[-\frac{La}{2},\frac{La}{2})^{N}$ respectively. Recall that $H_{L} - (-\frac{1}{2} \Delta_{L})$ is $-\frac{1}{2}\Delta_{L}$-bounded with zero relative bound. By the variational principle combined with the min-max principle, see e.g. \cite[Ex. 5.2.26]{BR2}, then $\forall \epsilon>0$ there exists a $C=C(\epsilon,N)>0$ s.t.
\begin{equation*}
(1 - \epsilon) \lambda_{k}^{(\mathrm{per})} - C \leq \nu_{k} \leq (1+\epsilon) \lambda_{k}^{(\mathrm{per})} + C,\quad k\in \mathbb{N}.
\end{equation*}
From the above inequality, we have for $k\neq 0$:
\begin{equation*}
\vert \frac{\nu_{k}}{\lambda_{k}^{(\mathrm{per})}} - 1 \vert \leq \epsilon + \frac{C}{\lambda_{k}^{(\mathrm{per})}},
\end{equation*}
and thus $\nu_{k} \sim \lambda_{k}^{(\mathrm{per})}$ when $k \rightarrow \infty$. Let us derive an asymptotic for the $\lambda_{k}^{(\mathrm{per})}$s. By Weyl's law:
\begin{equation*}
\lambda_{k}^{(\mathrm{Dir})} \sim \frac{4 \pi^{2}}{(\mathrm{V}_{N}(1) L^{N})^{\frac{2}{N}}}k^{\frac{2}{N}} \quad \textrm{when $k \rightarrow \infty$},
\end{equation*}
where $\mathrm{V}_{N}(1):=\pi^{\frac{N}{2}}/\Gamma(\frac{N}{2}+1)$ is the volume of the unit ball in $\mathbb{R}^{N}$; $\Gamma$ being the Gamma function. Since this asymptotic also holds true for the $\lambda_{k}^{(\mathrm{Neu})}$s, then \eqref{asymptotf} follows by a Neumann-Dirichlet bracketing-like argument. \qed \\

\noindent \textbf{Proof of Lemma \ref{trass}.} Let $L \in \mathbb{N}^{*}$ and $\beta>0$. Let $\{\mu_{k}\}_{k \geq 0}$, with $\mu_{k}=\mu_{k}(L,N_{L})$ be the set of eigenvalues of $H_{L,-}$ counting multiplicities and in increasing order. By the spectral theorem:
\begin{equation*}
\Vert H_{L,-}^{l} \mathrm{e}^{-\beta H_{L,-}} \Vert_{1} = \sum_{j=0}^{\infty} \vert \mu_{j}\vert^{l} \mathrm{e}^{-\beta \mu_{j}},\quad l \in \mathbb{N}^{*}.
\end{equation*}
To see that the above series is convergent, use the asymptotic in \eqref{asymptotf} along with the inequality:
\begin{equation*}
\forall x>0,\quad x^{\tau}\mathrm{e}^{-\sigma x} \leq (2\tau \mathrm{e}^{-1}\sigma^{-1})^{\tau} \mathrm{e}^{-\frac{\sigma}{2} x},\quad \tau,\sigma>0.  \tag*{\qed}
\end{equation*}

\noindent \textbf{Proof of Lemma \ref{relbord}}. We start by showing that under our assumptions, $P_{L,-}$ is $H_{L,-}$-bounded with zero relative bound. By the Cauchy-Schwartz inequality, we have $\forall \varphi \in D(H_{L,-})$:
\begin{equation}
\label{relatib00}
\Vert P_{L,-} \varphi \Vert_{2}^{2} \leq 2 N \langle \varphi, -\frac{1}{2}\Delta_{L,-} \varphi \rangle \leq 2 N \langle \varphi, H_{L,-} \varphi \rangle + 2N \vert \langle \varphi, \sum_{j=1}^{N} v_{\mathrm{per}} \varphi \rangle \vert,
\end{equation}
where we used in the last inequality that $v_{L,r_{0}} \geq 0$, see Lemma \ref{propvL}. Since $v_{\mathrm{per}}$ is $-\Delta$-bounded with zero-relative bound, then for any $\vartheta>0$ there exists $C(\vartheta)\in \mathbb{R}$ s.t.
\begin{equation*}
\vert \langle \varphi, \sum_{j=1}^{N} v_{\mathrm{per}} \varphi \rangle \vert \leq 2 \vartheta N \langle \varphi,-\frac{1}{2} \Delta_{L,-} \varphi \rangle + C(\vartheta) N \langle \varphi,\varphi \rangle.
\end{equation*}
Since $\vartheta$ can be chosen as small as we like, then there exist $0< c_{1}<1$ and $c_{2}(N) \in \mathbb{R}$ s.t.
\begin{equation}
\label{rzfst}
\vert \langle \varphi, \sum_{j=1}^{N} v_{\mathrm{per}} \varphi \rangle \vert \leq c_{1} \langle \varphi,H_{L,-} \varphi \rangle + c_{2}(N)  \langle \varphi,\varphi \rangle.
\end{equation}
Gathering \eqref{relatib00} and \eqref{rzfst} together, we arrive at:
\begin{equation}
\label{fdstr}
\Vert P_{L,-} \varphi \Vert_{2}^{2} \leq  2 N(1+c_{1}) \langle \varphi, H_{L,-} \varphi \rangle + 2N c_{2}(N)\langle \varphi,\varphi \rangle.
\end{equation}
By using the Cauchy-Schwartz inequality again, we then obtain:
\begin{equation}
\label{relatib1}
\forall \epsilon>0,\quad \Vert P_{L,-} \varphi \Vert_{2}^{2} \leq \epsilon^{2} \Vert H_{L,-} \varphi\Vert_{2}^{2}  + 2N (c_{2}(N) + \frac{1}{\epsilon^{2}})\Vert \varphi \Vert_{2}^{2}.
\end{equation}
$\mathrm{(i)}$ directly follows from \eqref{relatib1} by setting $\varphi = (H_{L,-} + i\alpha)^{-1} \psi$ with $\psi \in \mathfrak{h}_{L,-}^{N}$. We turn to $\mathrm{(ii)}$. Let $\zeta < 0$ s.t. $\zeta < \inf \sigma(H_{L,-})$. Such a choice is possible since $H_{L,-}$ is bounded from below. Write:
\begin{multline*}
P_{L,-} \mathrm{e}^{- i t H_{L,-}} P_{L,-} (H_{L,-}+i\alpha)^{-1} = \{P_{L,-} (H_{L,-} - \zeta)^{-\frac{1}{2}}\} \mathrm{e}^{- i t H_{L,-}} \{(H_{L,-} - \zeta)^{\frac{1}{2}} (-\frac{1}{2} \Delta_{L,-} + 1)^{-\frac{1}{2}}\} \\
 \times \{P_{L,-} (-\frac{1}{2} \Delta_{L,-} + 1)^{-\frac{1}{2}}\} \{(-\frac{1}{2} \Delta_{L,-} + 1)(H_{L,-}+i\alpha)^{-1}\}.
\end{multline*}
We now prove that each one of the operators between braces is bounded. From the first inequality in \eqref{relatib00} and \eqref{fdstr} respectively, there exists $C(N)>0$ s.t.
\begin{equation*}
\Vert P_{L,-} (-\frac{1}{2} \Delta_{L,-} +  1)^{-\frac{1}{2}}\Vert \leq 2 \sqrt{N},\quad
\Vert P_{L,-} (H_{L,-} - \zeta)^{-\frac{1}{2}}\Vert \leq C(N)(1 + \sqrt{-\zeta} + (\mathrm{dist}(\zeta,\sigma(H_{L,-})))^{-\frac{1}{2}}).
\end{equation*}
Besides, for any $\psi \in \mathfrak{h}_{L,-}^{N}$:
\begin{multline*}
\Vert (H_{L,-} - \zeta)^{\frac{1}{2}}(-\frac{1}{2}\Delta_{L,-}+1)^{-\frac{1}{2}} \psi \Vert_{2}^{2} \\
= \langle \psi, (-\frac{1}{2}\Delta_{L,-}+1)^{-\frac{1}{2}}[(-\frac{1}{2}\Delta_{L,-} + 1) + (H_{L,-}- (-\frac{1}{2}\Delta_{L,-})) - \zeta -1] (-\frac{1}{2}\Delta_{L,-}+1)^{-\frac{1}{2}} \psi \rangle.
\end{multline*}
Since $H_{L,-}- (-\frac{1}{2}\Delta_{L,-})$ is $-\frac{1}{2}\Delta_{L,-}$-bounded with zero relative bound, then there exists another $C(N)>0$ s.t. $\Vert (-\frac{1}{2}\Delta_{L,-}+1)^{-\frac{1}{2}}(H_{L,-}- (-\frac{1}{2}\Delta_{L,-}))(-\frac{1}{2}\Delta_{L,-}+1)^{-\frac{1}{2}}\Vert \leq C(N)$, see e.g. \cite[pp. 169]{RS2}. Therefore:
\begin{equation*}
\Vert (H_{L,-} - \zeta)^{\frac{1}{2}}(-\frac{1}{2}\Delta_{L,-}+1)^{-\frac{1}{2}} \psi \Vert_{2}^{2} \leq (2 + (-\zeta)  + C(N))\Vert \psi\Vert_{2}^{2}.
\end{equation*}
For the last operator between braces use that:
\begin{equation*}
\Vert (-\frac{1}{2} \Delta_{L,-} + 1)(H_{L,-}+i\alpha)^{-1} \Vert \leq (2+\frac{1}{\vert \alpha\vert}) + \Vert (H_{L,-} - (-\frac{1}{2} \Delta_{L,-})) (H_{L,-} + i\alpha)^{-1}\Vert.
\end{equation*}
Since $H_{L,-} - (-\frac{1}{2}\Delta_{L,-})$ is $-\frac{1}{2}\Delta_{L,-}$-bounded with zero relative bound, then it is $H_{L,-}$-bounded with zero relative bound. Ergo, the second term on the above r.h.s. is bounded by a constant $C(N)>0$.  \qed \\

\noindent \textbf{Proof of Lemma \ref{estimnorm1}.} $\forall L \in \mathbb{N}^{*}$ and $\forall T \geq  0$, the family $\{W_{L,-}(t),\,t \in (-\infty,T]\}$ is $H_{L,-}$-bounded with zero relative bound. Let us make it precise. From \eqref{defW} and by using \eqref{relatib1}, there exist two constants $c_{3}=c_{3}(\omega,\eta,T)>0$ and $c_{4}=c_{4}(\omega,\eta,T)>0$ s.t. $\forall t \in (-\infty,T]$ and $\forall E \in [-1,1]$:
\begin{equation}
\label{relatib2}
\forall \epsilon>0,\quad \Vert W_{L,-}(t) \varphi \Vert_{2} \leq c_{3} \epsilon \Vert H_{L,-} \varphi \Vert_{2} + \{c_{3} C(N) (1 + \frac{1}{\epsilon}) + c_{4}N\}\Vert \varphi \Vert_{2},
\end{equation}
which holds $\forall \varphi \in D(H_{L,-})$. By taking $\varphi = (H_{L,-}+i\alpha)^{-1}\psi$, with $\psi \in \mathfrak{h}_{L,-}^{N}$ and $\alpha \in \mathbb{R}^{*}$:
\begin{equation}
\label{estirel}
\sup_{t \in (-\infty,T]} \Vert W_{L,-}(t) (H_{L,-}+i\alpha)^{-1}\Vert \leq  c_{3} \epsilon   + \{c_{3} C(N) (1 + \frac{1}{\epsilon}) + c_{4}N\}\frac{1}{\vert \alpha\vert}.
\end{equation}
We turn to $\mathrm{(ii)}$. Fix $\epsilon$ in \eqref{estirel}, say $\epsilon^{-1} =  4\max\{c_{3},c_{4}\}$. Pick $\gamma_{0} \in \mathbb{R}^{*}$ with $\vert \gamma_{0}\vert$ large enough s.t.:
\begin{equation*}
2 \sup_{t \in (-\infty,T]} \Vert W_{L,-}(t) (H_{L,-}+i\gamma_{0})^{-1}\Vert \leq 1.
\end{equation*}
Note that $\gamma_{0} = \gamma_{0}(N,\omega,\eta,T)$. By iterating the second resolvent equation, we infer the  estimate:
\begin{equation*}
\sup_{t \in (-\infty,T]}\Vert (H_{L,-}+i\gamma_{0})(H_{L,-}(t) + i\gamma_{0})^{-1}\Vert \leq 3.
\end{equation*}
To obtain $\mathrm{(ii)}$ it remains to use the following identities (below, $\alpha \in \mathbb{R}^{*}$):
\begin{gather*}
(H_{L,-}+i\alpha)(H_{L,-}(t)+i\alpha)^{-1} = (H_{L,-}+i\gamma_{0}) (H_{L,-}(t)+i\alpha)^{-1} + i(\alpha-\gamma_{0})(H_{L,-}(t)+i\alpha)^{-1},\\
(H_{L,-}+i\gamma_{0}) (H_{L,-}(t)+i\alpha)^{-1} = (H_{L,-}+i\gamma_{0}) (H_{L,-}(t)+i\gamma_{0})^{-1}[\mathbbm{1} + i(\alpha-\gamma_{0}) (H_{L,-}(t)+i\alpha)^{-1}]. \tag*{\qed}
\end{gather*}

\section{Acknowledgments.}

The three authors warmly thank Horia D. Cornean for many fruitful discussions. B.S. thanks Tony C. Dorlas for stimulating discussions related to this topic.

\end{document}